\documentclass[a4paper,fleqn,usenatbib, useAMS]{mnras}
\usepackage{newtxtext,newtxmath}
\usepackage[T1]{fontenc}
\usepackage{ae,aecompl}

\usepackage{graphicx}	
\usepackage{amsmath}	
\usepackage{amssymb}	
\usepackage{multicol}        
\usepackage{bm}		
\usepackage{pdflscape}	
\usepackage{tabularx}

\newcommand{\vect}{\ensuremath{\mathbf}}
\newcommand{\dropsign}[1]{\smash{\llap{\raisebox{-.5\normalbaselineskip}{$#1$\hspace{2\arraycolsep}}}}}%
\newcommand*\samethanks[1][\value{footnote}]{\footnotemark[#1]}
\newcommand{\ms}{\,m\,s$^{-1}$} 
\newcommand{\mms}{\,m$^{2}$\,s$^{-1}$} 

\title[Method for solving problems of CR modulation]{An analytically iterative method for solving problems of cosmic-ray modulation}

\author[Yu. L. Kolesnyk et al.]{
Yuriy L. Kolesnyk,$^{1}$\thanks{E-mail: \href{mailto:kolesnyk@mao.kiev.ua}{kolesnyk@mao.kiev.ua} (YK); \href{mailto:bobik@saske.sk}{bobik@saske.sk} (PB)}
Pavol Bobik,$^{2}$\samethanks
Boris A. Shakhov,$^{1}$
Marian Putis$^{2}$
\\
$^{1}$Main Astronomical Observatory, National Academy of Sciences of Ukraine, 27, Akademika Zabolotnoho St., UA-03680 Kyiv, Ukraine\\
$^{2}$Institute of Experimental Physics, Watsonova 47, 04001 Kosice, Slovakia\\
}
\date{Accepted 2017 May 11. Received 2017 May 11; in original form 2017 February 13}

\pubyear{2017}
\begin{document}
\label{firstpage}
\maketitle
\parindent=0.3cm
\voffset=-1.2cm
\textheight=245mm 
\begin{abstract}
The development of an analytically iterative method for solving steady-state as well as unsteady-state problems of cosmic-ray (CR) modulation is proposed. Iterations for obtaining the solutions are constructed for the spherically symmetric form of the CR propagation equation. The main solution of the considered problem consists of the zero-order solution that is obtained during the initial iteration and amendments that may be obtained by subsequent iterations. The finding of the zero-order solution is based on the CR isotropy during propagation in the space, whereas the anisotropy is taken into account when finding the next amendments. To begin with, the method is applied to solve the problem of CR modulation where the diffusion coefficient $\kappa$ and the solar wind speed $u$ are constants with an Local Interstellar Spectra (LIS) spectrum. The solution obtained with two iterations was compared with an analytical solution and with numerical solutions. Finally, solutions that have only one iteration for two problems of CR modulation with $u = $~constant and the same form of LIS spectrum were obtained and tested against numerical solutions. For the first problem, $\kappa$ is proportional to the momentum of the particle $p$, so it has the form $\kappa=k_0\eta$, where $\eta=\frac{p}{m_0c}$. For the second problem, the diffusion coefficient is given in the form $\kappa=k_0\beta\eta$, where $\beta=\frac{\upupsilon}{c}$ is the particle speed relative to the speed of light. There was a good matching of the obtained solutions with the numerical solutions as well as with the analytical solution for the problem where $\kappa = $~constant.
\end{abstract}
\section{Introduction}
The classic form of the equation of transport for cosmic rays (TPE) in a spherically symmetric interplanetary region, without taking into account particle drift and any local sources, is \citep{Parker, Dolginov1967, GleesonAxford1967, Dolginov1968, JokipiiParker}
\begin{equation}
\label{eq. TPE}
\frac{1}{r^2}\frac{\upartial}{\upartial r}\left (r^2\kappa\frac{\upartial f}{\upartial r} \right) - u\frac{\upartial f}{\upartial r} + \frac{1}{3r^2}\frac{\upartial }{\upartial r} (r^2u)\frac{\upartial f}{\upartial \ln{p}}  = \frac{\upartial f}{\upartial t},
\end{equation}
where $f(\mathbfit{r},\:p,\:t)$ is the omnidirectional distribution function, $\mathbfit{r}$ is the radial distance from the Sun, $p$ is the modulus of the momentum of the particle, $t$ is time, $u$ is the radial solar wind (SW) speed and $\kappa(r, p, t)$ is the effective radial diffusion coefficient, i.e. it is $\kappa_{rr}$, which is the one from the nine elements of the diffusion tensor $\mathbfit{K}$.

The omnidirectional distribution function, $f(\mathbfit {r},\:p,\:t)$, relates to the full cosmic-ray (CR) distribution function, $F(\mathbfit{r}, \:\mathbfit{p},\:t)$, as 
\begin{equation}
\label{eq. f}
f(\mathbfit{r},\:p,\:t)=\frac{1}{4\upi}\int F(\mathbfit{r},\:\mathbfit{p},\:t) \mathrm{d}\Omega \:.
\end{equation}
On the other hand, if particles are scattered by random inhomogeneities so that the angular distribution of the directions of the particles becomes isotropic, then it is possible to apply the diffusion approximation \citep{Morse1953}, i.e. to use the moments of $F(\mathbfit{r},\:\mathbfit{p},\:t)$:
\begin{equation}
\label{eq. N}
N(\mathbfit{r},\:p,\:t)=\int F(\mathbfit{r},\:\mathbfit{p},\:t) \mathrm{d}\Omega \:,
\end{equation}
\begin{equation}
\label{eq. J}
\mathbfit{J}(\mathbfit{r},\:p,\:t)=\int F(\mathbfit{r},\:\mathbfit{p},\:t) \bm{\upupsilon} \mathrm{d}\Omega \:,
\end{equation}
where $N(\mathbfit{r},\:p,\:t)$ is the phase density of the particles and $\mathbfit{J}(\mathbfit{r},\:p,\:t)$ is the vector of the particle flux in space. The integration in (\ref{eq. f}), (\ref{eq. N}) and (\ref{eq. J}) is over the solid angle element $\mathrm{d}\Omega$ in momentum space.
Thus, taking into account (2) and (3), the TPE may be rewritten, as in \citet{Dolginov1967, Dolginov1968}, in a form involving $N(\mathbfit{r}, p, t)$. Note that  in the form was obtained by \citet{Dolginov1967, Dolginov1968} from the kinetic Boltzman equation in which the regular as well as the random components of the interplanetary magnetic fields were taken into account. Such form can be represented as a continuity equation that describes the propagation of particles for a spherically symmetric event in phase space \citep{Dorman1978, GleesonWebb}:
\begin{equation}
\label{eq. s.s. event}
\nabla_r{j_r}+\nabla_p{j_p} = -\frac{\upartial N(r,p,t)}{\upartial t}\:,
\end{equation}
where
$\nabla_r = \frac{1}{r^2}\frac{\upartial}{\upartial r}r^2$ is the radial component of the divergence operator and $\nabla_p = \frac{1}{p^2}\frac{\upartial}{\upartial p}p^2$ is the component of the divergence operator in momentum space. The value
\begin{equation}
\label{flow r}
{j_r} = - \kappa(r,\:p,\:t)\frac{\upartial N(r,\:p,\:t)}{\upartial r}-\frac{up}{3}\frac{\upartial N(r,\:p,\:t)}{\upartial p}
\end{equation}
is the CR flux in space, and
\begin{equation}
\label{flow p}
j_p = \frac{up}{3}\frac{\upartial N(r,\:p,\:t)}{\upartial r}
\end{equation}
is the CR flux in momentum space. For stationary case, equation (5) describes CR propagation that is reduced to CR advection by the SW. Particles with different energies are transferred with different rates, which is described by term $-\frac{up}{3}\frac{\upartial N(r,\:p,\:t)}{\upartial p}$, and, on the other hand, CRs penetrate diffusively into the SW with the rate $\kappa(r,\:p,\:t)\frac{\upartial N(r,\:p,\:t)}{\upartial r}$. This is 
accompanied by a change in the particle energy, which is determined by the balance between the oncoming and following encounters of particles with the moving irregularities of the interplanetary magnetic field ($\frac{up}{3}\frac{\upartial N(r,\:p,\:t)}{\upartial r}$). 

The possibility to obtain the exact analytical solutions (\ref{eq. TPE}) for various forms of $\kappa$ and $u$ [not included when (\ref{eq. TPE}) is greatly simplified for $u\propto\frac{1}{r^2}$] has always attracted the attention of researchers. In this way, significant results were obtained, but only for the steady-state case with $u=const$ and involuntary forms of $\kappa$, i.e. a form of $\kappa$ that does not depend simultaneously on $r$, $p$ or $R$ (the particle rigidity) and $\beta$, where $\beta=\frac{\upupsilon}{c}$ is the particle speed relative to the speed of light. In particular, analytic solutions have been derived by \citet{Dolginov1967, Dolginov1968} for $\kappa=$~constant (see Appendix A, equation \ref{N_ex.sol.}), by \citet{FiskAxford1969} for $\kappa=k_0T^{\alpha}r^b$ ($b>1$, T is particle kinetic energy), by \citet{Webb1977} for $k_0(p)r^b$ (via Green's theorem and Green's function) and by \citet{Shakhov2008} for $k_0r^b$ (via Mellin transform). Attempts to obtain the exact analytical solution (\ref{eq. TPE}) for arbitrary forms of  $\kappa$ and $u$ and, especially, for unsteady cases encounter serious mathematical difficulties. Therefore, the development of the approximate analytical methods but with an acceptable accuracy for solving steady-state as well as time-dependent problems is an urgent problem. 
Furthermore, often it is necessary to have the qualitative characteristics of the processes and the ability to freely operate by the variables in analytical formulae. This is a big advantage compared to the numerical calculations, which, moreover, are almost always time-consuming.
\section{About the method}
In this paper, we continue with the development of an analytically iterative method for solving problems of CR modulation. The method was initially proposed by \citet{Shakhov2006, Shakhov2009}. The point of the method is as follows: Let us represent the particle density as the sum
\begin{equation}
\label{common solution}
N=N_{0}(r,\:p,\:t)+N_{1}(r,\:p,\:t)+N_{2}(r,\:p,\:t)+...
\end{equation}
where $N_{0}(r,\:p,\:t)$ is the zero-order solution and $N_{1,2,3..}(r,\:p,\:t)$ are amendments to the zero solution. To obtain the zero-order solution, we assume that the anisotropy of the CRs during propagation in space is small, and that there is no radial CR flux at any point of space. This is the force-field approximation, which was initially introduced by \citet{GleesonAxford1968} and furthered in \citep{GleesonUrch}, which presented a new development for its solutions.
At the boundary of the CR modulation ($r_{0}$), for example, at the heliosphere boundary, the zero-order solution must satisfy a boundary condition involving its spectrum; for example be equal to some LIS spectrum. So, to find of the zero-order solution, the following conditions must be satisfied:
\begin{equation}
\label{first condition}
\begin{cases}
j_{r,0}(r,\:p,\:t)=0,\\
N_{0}(r_{0},\:p,\:t)=\mathrm{LIS}.\
\end{cases}
\end{equation}
The amendments to the zero-order solution can be obtained from~(\ref{eq. s.s. event}) by the following recurrence relation:
\begin{equation}
\nabla_r{j_{r,i+1}}+\nabla_p{j_{p,i}} = -\frac{\upartial N_i(r,p,t)}{\upartial t},
\end{equation}
where $i$ is an iteration index, which changes from the zero to infinity.

Taking into account~(\ref{flow r}) and~(\ref{flow p}), the last relation can be written in the following form:
\begin{align}
\label{iteration eq}
\frac{1}{r^2}&\frac{\upartial}{\upartial r}r^2\left[-\kappa\frac{\upartial N_{i+1}}{\upartial r}-\frac{up}{3}\frac{\upartial N_{i+1}}{\upartial p}\right]+\frac{1}{p^2}\frac{\upartial}{\upartial p}p^2\left[\frac{up}{3}\frac{\upartial N_i}{\upartial r}\right]\nonumber \\
&=-\frac{\upartial N_i}{\upartial t}.
\end{align}
Here, in order to obtain $N_{i+1}$, we will be guided by the fact that any radial CR flux (such as from the zero solution and also from the amendments) in the centre of the heliosphere (near the Sun) must be absent, i.e. $j_{r,i}(0,\:p,\:t)=0$. This follows from the spherical symmetry of the problem. It should be noted that the amendments of the fluxes will be small but not equal to zero. This means that as more amendments are obtained, more of the CR anisotropy in the space will be taken into account. And the last condition is for finding $N_{i+1}$. As a common solution is presented in the form~(\ref{common solution}) and the zero solution $N_0$ satisfies the boundary condition at $r_0$, this means that all amendments $N_{i+1}$ at $r_0$ must be equal to zero.

To sum up the above, for the finding of the amendments necessary to fulfil such conditions,
\begin{equation}
\label{second condition}
\begin{cases}
j_{r,i+1}(0,\:p,\:t)=0,\\
N_{i+1}(r_{0},\:p,\:t)=0.\
\end{cases}
\end{equation}
Thus, in the first step, using~(\ref{flow r}) and~(\ref{first condition}), we may obtain the zero solution [$N_0(r,\:p,\:t)$]. Then, inserting it into~(\ref{iteration eq}) and applying~(\ref{second condition}), we obtain the first amendment [$N_1(r,p,t)$]. For obtaining the next amendment, it is necessary to substitute $N_1(r,\:p,\:t)$ into~(\ref{iteration eq}) and again apply~(\ref{second condition}), and so on. A common solution, as mentioned above, will be expressed in the form~(\ref{common solution}).

In the following sections of this paper, we test the analytically iterative method (AI) for a constant diffusion coefficient and two shapes of non-constant diffusion coefficient.

We used the stochastic integration B-p method and the Crank-Nicolson method (CN) to verify the solutions obtained by AI.
B-p is a stochastic integration method to solve the TPE backwards in time. To find stochastic solutions of the Fokker-Planck equation (FPE), Itos lemma was used to rewrite the TPE into a set of stochastic differential equations. This method historically is treated as a forward-in-time method of solution. The backward-in-time solution was introduced by \citet{Kota1977} and was described in detail in, for example, \citet{Zhang}. We apply the 1D backward stochastic integration method that is used, for example, by \citet{Yamada}. A comparison of the backward and forward methods with a systematic error estimation for realistic LIS was published in \citet{Bobik}, where the B-p method is described in detail.

CN is a numerical unconditionally stable finite-difference method to solve FPEs \citep{Diaz}. The method was proposed by \citet{Fisk1971} to solve TPEs, and then used by many other authors. Here, the CN method is applied in the form  given in \citet{Batalha2013}. 

To compare the AI method, we use two completely independent numerical methods, to avoid any numerical artefacts or errors in the verification of the method.
\begin{figure}
	\centering
	\includegraphics[width=.36\textwidth]{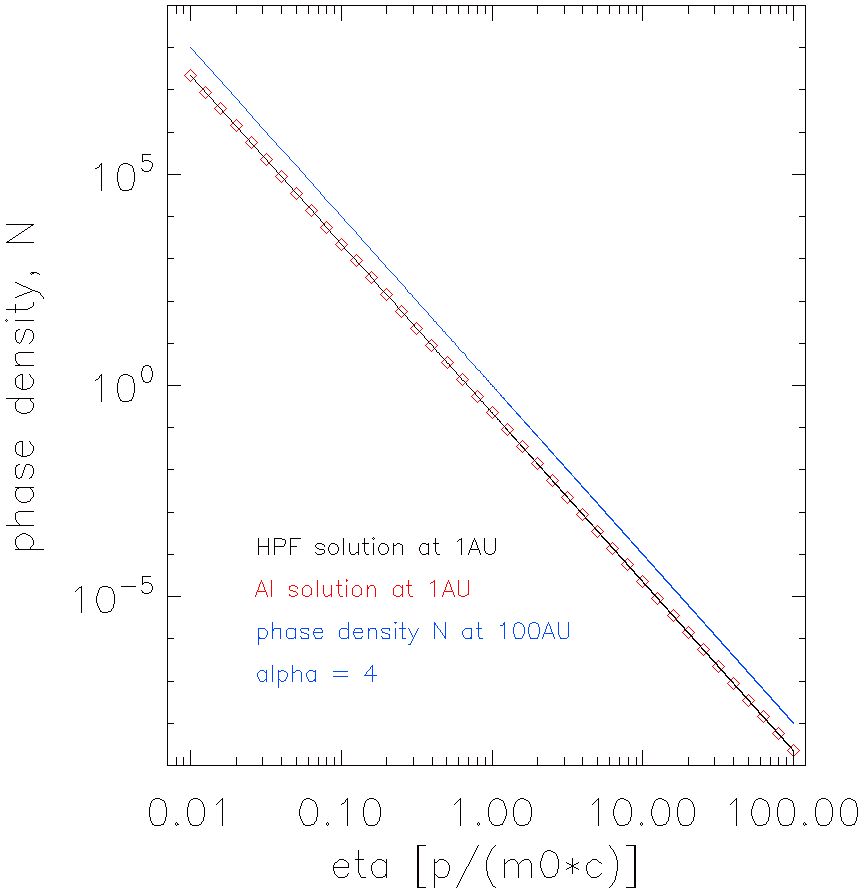}
	\caption{The comparison of the AI and HPF solutions at 1 au for the constant diffusion coefficient $\kappa = 5\times10^{18}$\mms\: and $\alpha=4$ (for more details, see the text)}
	\label{fig1}
\end{figure}
\section{A constant diffusion coefficient}
To verify the AI approach, we start with the simpler case where the diffusion coefficient $\kappa$ is constant.
Obtaining a solution by the AI method for $\kappa = $~constant and an LIS at the heliosphere boundary that has the form $Cp^{-\alpha}$ is discussed in Appendix A.
A comparison with an analytical solution (see equation \ref{N_ex.sol.}) that is based on the confluent hypergeometric functions (HPF method) is available for this case, which therefore allows us to very accurately estimate the precision of the AI method.
If not mentioned otherwise, all results published in this paper were evaluated for protons in a spherically symmetric heliosphere with radius 100 au, where the SW has a speed of 400 km s$^{-1}$.

In Fig.~\ref{fig1}, the phase density function $N$ at $1$ au is shown for $\kappa = 5\times10^{18}$\mms\: as are the slopes of the initial spectrum at the heliosphere boundary $\alpha = 4$
evaluated by AI (red diamonds) and by HPF (black line), together with $N$ at the modulation boundary (blue line). Both solutions are very similar at $1$ au.
The difference between them is of the order of 0.01 per cent at all energies from $\eta=0.01$ ($\eta= \frac{p}{m_0 c}$, a proton with kinetic energy 47 keV) to $\eta=100$ (a proton with kinetic energy 92.89 GeV).
As can be seen from  Fig.~\ref{fig1}, the modulation is relatively the same for all energies because the diffusion coefficient does not depend on the energy. More precisely, by saying modulation, we mean the ratio between the phase density inside the heliosphere (here at 1 au) and that normalised to the phase density for the same momentum at the modulation boundary (at 100 au). From Fig.~\ref{fig1}, we can see that difference between the solution that was obtained by AI and that obtained by HPF at the selected $r$ does not depend on the energy.

We present in Fig.~\ref{fig2} a comparison of the phase density spatial distribution across the heliosphere for selected energies, as evaluated by the AI, HPF, and CN methods.
\begin{figure}
\label{fig2}
	\centering
	\includegraphics[width=0.47\textwidth]{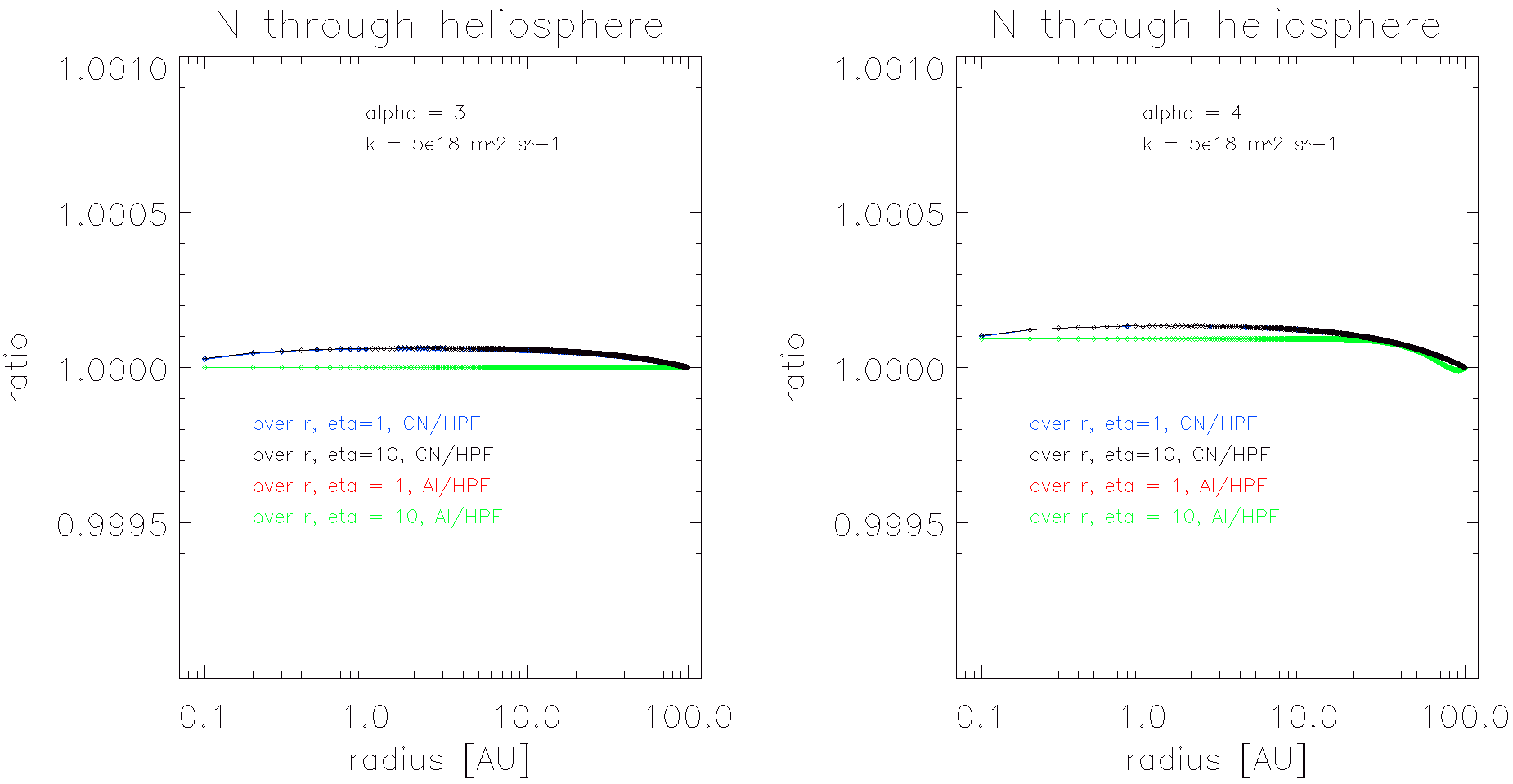}
	\caption{The comparison of the ratios of the AI, HPF and CN solutions as functions of the radius for the case of a constant diffusion coefficient $\kappa = 5\times10^{18}$\mms\: and $\alpha=3$ (left-hand panel) and $\alpha=4$ (right-hand panel) (for more details, see the text)}
	\label{fig2}
\end{figure}
\begin{figure}
	\centering
	\includegraphics[width=.47\textwidth]{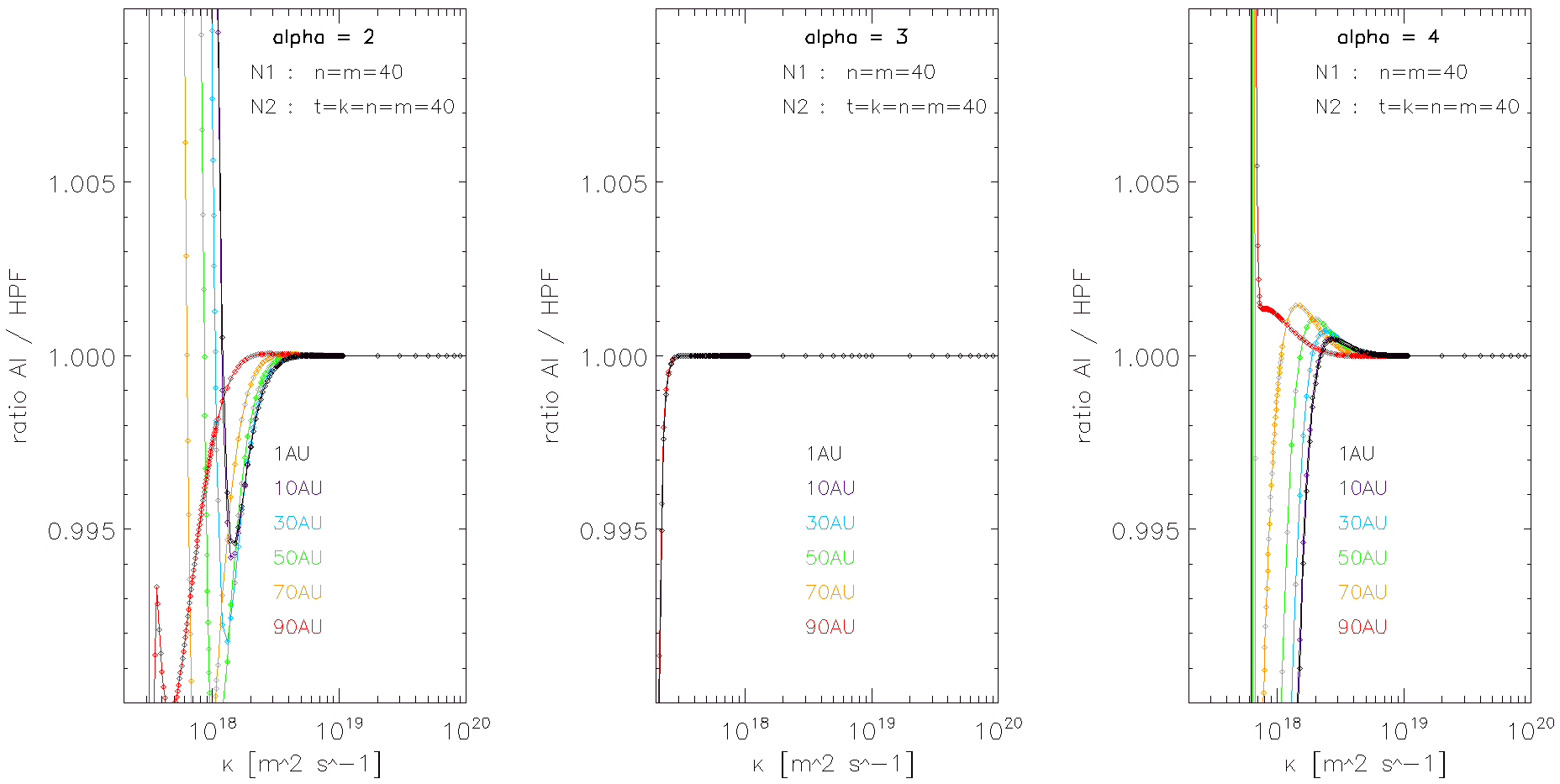}
	\caption{The comparison of the ratios of the AI and HPF solutions as functions of the value of a constant diffusion coefficient for $\alpha=2$ (left-hand panel), 3 (right-hand panel) and 4 (right panel) (for more details, see the text)}
	\label{fig3}
\end{figure}
Presented are the ratios of the phase-density functions evaluated for $\kappa = 5\times10^{18}$\mms\: and the slopes of the initial spectra at the heliosphere boundary $\alpha = 3$
(left-hand panel) and $\alpha = 4$ (right-hand panel). As for the case of constant $\kappa$, the ratio of AI/HPF does not depend on the energy (as can be seen from the AI solution and the exact solution \ref{N_ex.sol.}); thus, the ratios as functions of the radius for $\eta=1$ (388 MeV) and $\eta=10$ (8.49 GeV) are very similar. Therefore, the green line in Fig.~\ref{fig2} representing the ratio for $\eta=10$ hides the red line of the $\eta=1$ solution in both figure panels.

For the special case when $\alpha=3$ with the mentioned value of $\kappa$, the AI solution is very precise, with a difference between the AI and HPF solutions at a level of $10^{-6}$ per cent.
This is due to the fact that for this case, the first amendment $N_1$ and the second amendment $N_2$ (as well as all higher amendments that will be found) are equal to zero (see Appendix A). As a result, a common solution via AI is described by $N_0$ alone (equation \ref{N_0_k_const}) and is $N_{0}=C\eta^{-3}\exp[x-x_{0}]$. Moreover, HPF describes equation~(\ref{N_ex.sol._alpha_3}) (see Appendix A) for this case. Therefore, HPF and AI yield identical solutions for $\alpha=3$. The range of numerical error during the evaluation of the solutions is then very small.
In Fig.~\ref{fig2}, we also show the ratios of the phase-density functions evaluated by CN and HPF. CN, in both cases, attains a precision of $0.01$ per cent (precision as the difference from the analytic HPF solution). You can see that the AI solution for $\alpha=4$ is more precise than the CN solution. We evaluated the solutions for the same set of parameters also by the stochastic B-p method. The results from the B-p method confirmed the results presented by the AI and HPF methods.

\begin{figure}
	\centering
	\includegraphics[width=.35\textwidth]{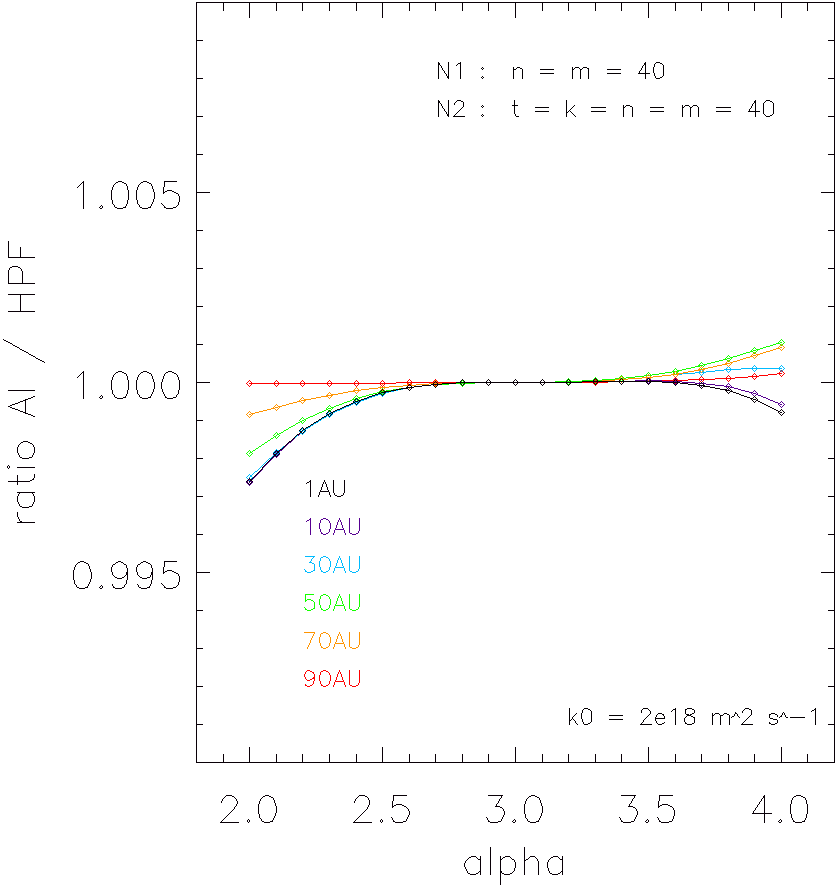}
	\caption{The comparison of the ratio of the AI and HPF solutions as functions of the initial spectrum slope $\alpha$ for a constant $\kappa$ (for more details, see the text)}
	\label{fig4}
\end{figure}

The precision of the AI solution depends on $\kappa$ and slightly less on $\alpha$.
In Fig.~\ref{fig3}, we present the dependence of the ratio between the AI and HPF solutions on $\kappa$ at different distances $r$.
For $\alpha=2$ and 4, the AI solutions are precise, of the order of $0.1$ per cent precision level, for diffusion coefficients $\kappa$ greater than $2\times10^{18}$\mms.
As could be expected, for $\alpha=3$, we see a very high precision at every position inside the heliosphere for $\kappa>3\times10^{17}$\mms.
For $\kappa=2\times10^{18}$\mms, we evaluated the ratio of the AI and HPF solutions in dependence on $\alpha$. The results are presented in Fig.~\ref{fig4}. The precision decreases as $\alpha$ becomes more different from $3$ and with decreasing $r$.

It should be noted that the precision of the AI solution may be improved by finding more amendments (as mentioned in the description of the method) and by increasing the order of the summations in the evaluation of the amendments. This is true in the presented case for the first and the second amendments. The latter is understandable, due to the fact that we have applied a Taylor series expansion for the exponential to obtain these amendments. But this has a limit due to limitations of the data types of computing languages. In particular, if a higher value of $\frac{ur}{\kappa}$ is used, then the number of terms in the sums in $N_1$ and $N_2$ must be increased in order to attain an acceptable accuracy. As a result, this situation significantly increases the time for the evaluation of the amendments.
\section{Non-constant diffusion coefficient}
\subsection{Solution for \mbox{\boldmath$\kappa = k_0\eta$}}
The solution for the case of a non-constant diffusion coefficient with shape ${\kappa = k_0\eta}$ is derived in Appendix B on the basis of the AI method. Here, it should be noted that although $N_1$ (see equation \ref{N_1_eta}) is expressed through integrals, but to construct the corresponding curve, it is sufficient to use one of the mathematical packages such as {\scriptsize MAPLE}, {\scriptsize MATHEMATICA}, {\scriptsize MATLAB}, etc. In this case, the result will be obtained faster than the numerical calculation of such a problem. At the same time, numerical calculations (for CN as well as the B-p stochastic method) of the problem should use or apply the program code to obtain the required points. And then based on these points and by means of the software, we can build the appropriate curve.

An analytical solution for this shape of the diffusion coefficient and power-law unmodulated spectrum is not available. In particular, it should be noted that for a similar problem where for $\kappa=k_0p^\alpha$, an analytical solution was obtained by \citet{Cowsik} only for the constant SW speed and for a specific form of LIS spectrum, but not for an arbitrary form. 
Herewith, the form of LIS spectrum that was examined is not a power-law kind.
\begin{figure}
	\centering
	\includegraphics[width=.47\textwidth]{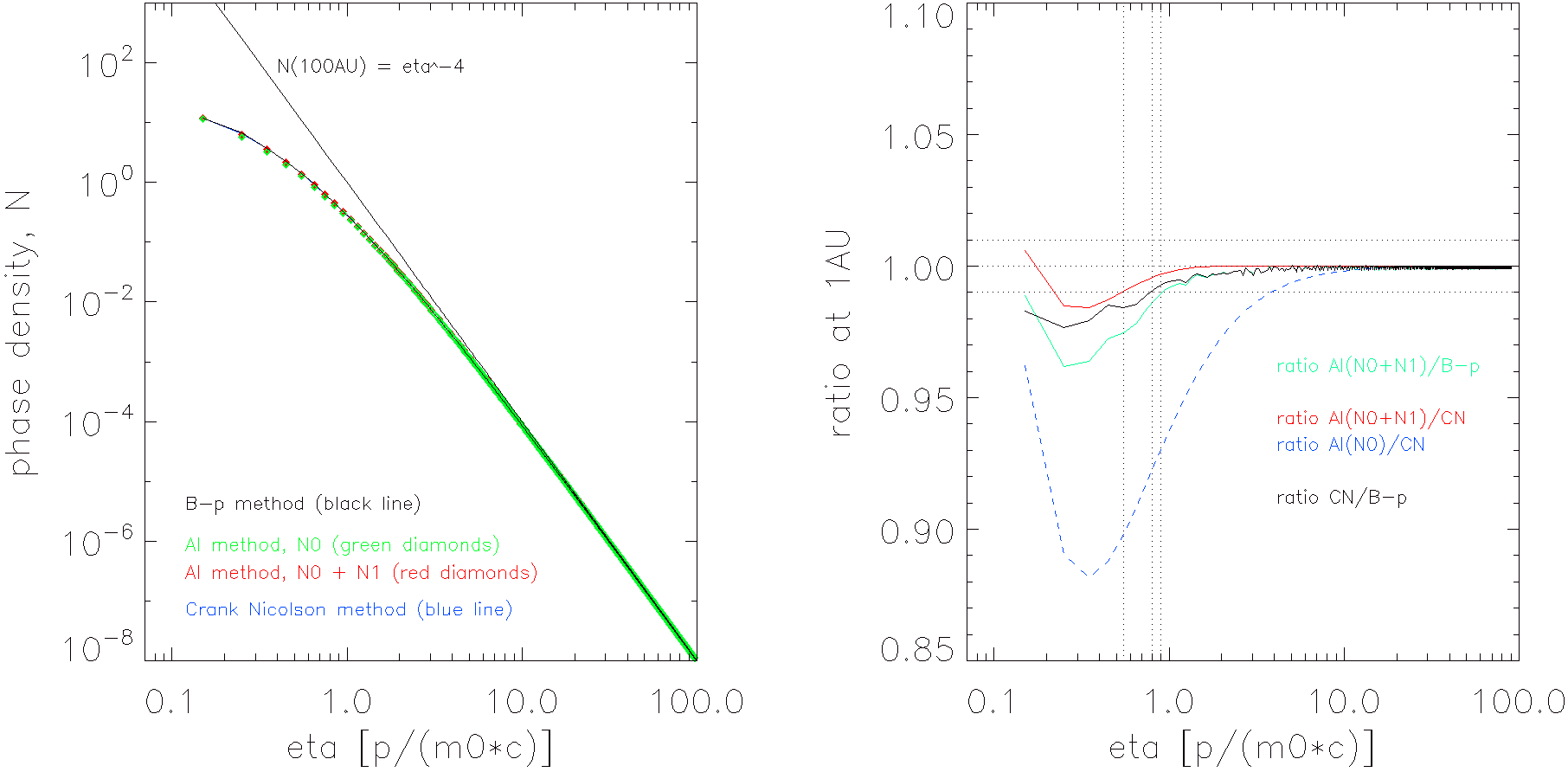}
	\caption{The comparisons of the AI, B-p and the CN solutions at 1 au for $k_0=5\times10^{18}$\mms, $\alpha=4.0$ and ${\kappa = k_0\eta}$ (for more details, see the text)}
	\label{fig5}
\end{figure}
\begin{figure}
	\centering
	\includegraphics[width=.47\textwidth]{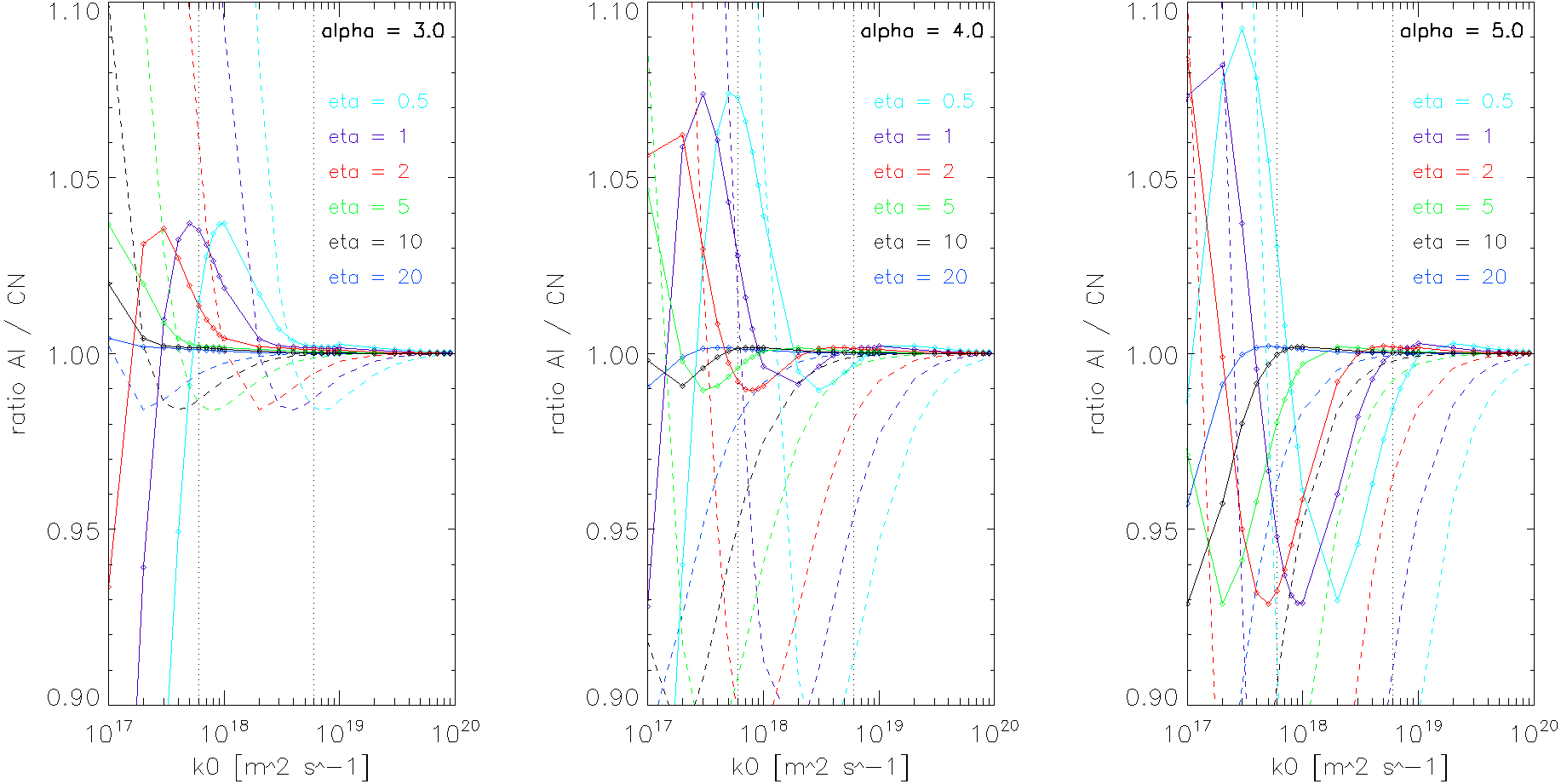}
	\caption{The comparison of AI and CN solutions ratios for different diffusion coefficients $k_0$ and $\alpha=3$ (left-hand panel), 4 (middle panel) and 5 (right-hand panel) at 1 au for $\eta=$~0.5, 1, 2, 5, 10 and 20. Dotted lines represent the ratio of AI$(N_0)$/CN solutions and solid lines represent the ratio of AI$(N_0+N_1)$/CN. AI and CN for ${\kappa=k_0\eta}$ (for more details, see the text)}
	\label{fig6}
\end{figure}

We compared the AI solutions, i.e. $N_0$ and $N_0+N_1$, with the solutions provided by CN and the B-p stochastic method. By using two independent numerical methods (B-p and CN), we want to ensure the validity of the comparison with the AI method.

In Fig.~\ref{fig5}, a comparison of the solutions for energies starting from $\eta=0.15$ ($T=11$~MeV) at 1 au for $\alpha=4.0$ and $k_0=5\times10^{18}$\mms is presented. The zero-order solution $N_0$ is indicated in the figure by AI$(N_0)$, and the sum of the zero-order solution and the first amendment $N_0+N_1$ is indicated by AI$(N_0+N_1)$.
Over a wide range of energies, all four solutions, i.e. $N_0$, $N_0+N_1$, CN and B-p, in Fig.~\ref{fig5} are similar. 
As can be seen from the right-hand panel of Fig.~\ref{fig5}, comparisons of the AI solution with that of the CN method, and of the AI solution with B-p, show that the difference is less than 1 per cent for energies higher than 324 MeV ($\eta>0.9$).
The CN and AI solutions are more similar than the AI and B-p solutions. The ratio AI/CN remains within the 1 per cent difference range for energies higher than 133 MeV ($\eta>0.55$). The CN solution and the B-p solution differ by less than 1 per cent over a wide range of energies: from 263 MeV ($\eta=0.8$) to higher energies.

To look more closely at the precision of the AI solution, we compare the solutions provided by the AI and CN methods at 1 au for a range of values of the diffusion coefficient and for the selected slopes of the power-law unmodulated spectrum at 100 au.
In Fig.~\ref{fig6}, the ratios between the AI and CN solutions at 1 au for different ${\kappa_0}$ and ${\alpha=3.0}$, $4.0$ and $5.0$ and selected energies are shown. The dotted lines represent the ratio of the solutions AI$(N_0)$/CN and the solid lines represent the ratio AI$(N_0+N_1)$/CN. Here, AI$(N_0)$ is the zero-order solution and AI$(N_0+N_1)$ is the sum of the zero-order solution and the first amendment.
The ratios become closer to unity with increasing values of
$\kappa_0$. We can see from the figure that when $k_0$ is larger than $2\times10^{17}$\mms, then all AI$(N_0+N_1)$ differ from the CN solutions by less than 1 per cent for energies over $\eta=20$ (17.85 GeV) and in the range of $\alpha$ tested. Also, for $k_0$ larger than $2\times10^{18}$\mms, all AI solutions differ from CN solutions by less than 1 per cent for energies over $\eta=2$ (1.16 GeV).  
\begin{figure}
	\centering
	\includegraphics[width=.47\textwidth]{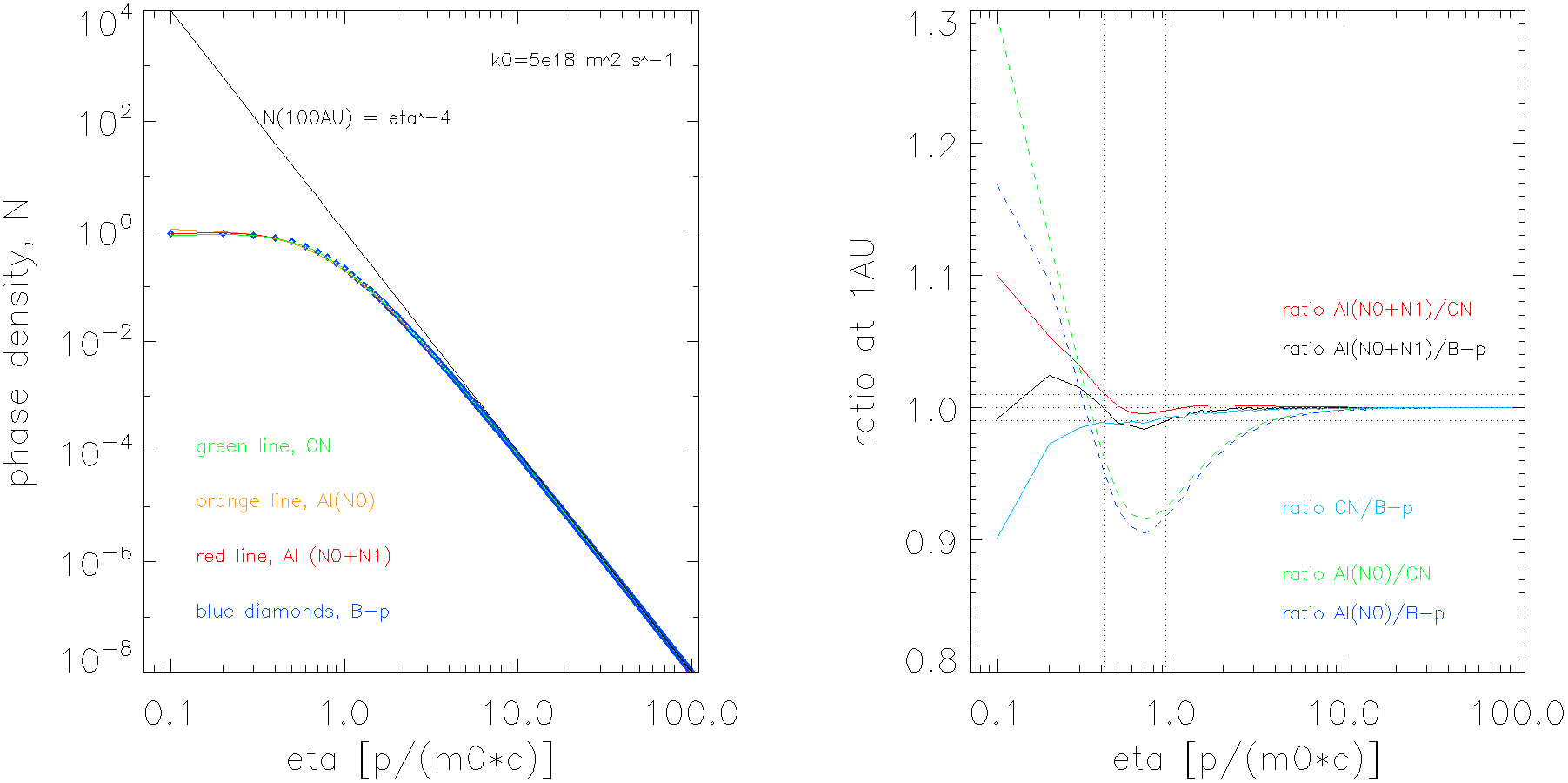}
	\caption{The comparison of the AI, B-p and CN solutions at 1 au for $k_0=5\times10^{18}$\mms, $\alpha=4.0$ and ${\kappa = k_0\beta\eta}$ (for more details, see the text)}
	\label{fig7}
\end{figure}
\subsection{Solution for \mbox{\boldmath$\kappa = k_0\beta\eta$}}
The solution for a non-constant diffusion coefficient with shape ${\kappa = k_0\beta\eta}$ and for a power-law unmodulated spectrum is derived in Appendix C.
The comparison of the AI solutions with the solutions that were obtained by the CN and B-p methods for energies starting from $\eta=0.1$ ($T=5$ MeV) at 1 au for $k_0 = 5\times10^{18}$\mms and $\alpha=4.0$ is presented in Fig.~\ref{fig7}. 

The ratios of the AI solution to CN and B-p solutions for the zero-order solution $N_0$ [indicated by AI$(N_0)$] and for the common solution that consists of the zero solution and the first amendment $N_0+N_1$ [indicated by AI$(N_0+N_1)$] are shown in the figure's right-hand panel. 

From the inspection of the figure, we can see that the common AI solution differs by less than 1 per cent from the CN solution for energies bigger than 79 MeV ($\eta>0.42$). The B-p solution stays within the 1 per cent difference region from energies $\oldge$ 349 MeV ($\eta>0.94$), but for smaller energies, it oscillates around the 2 per cent difference region.

\begin{figure}
	\centering
	\includegraphics[width=.47\textwidth]{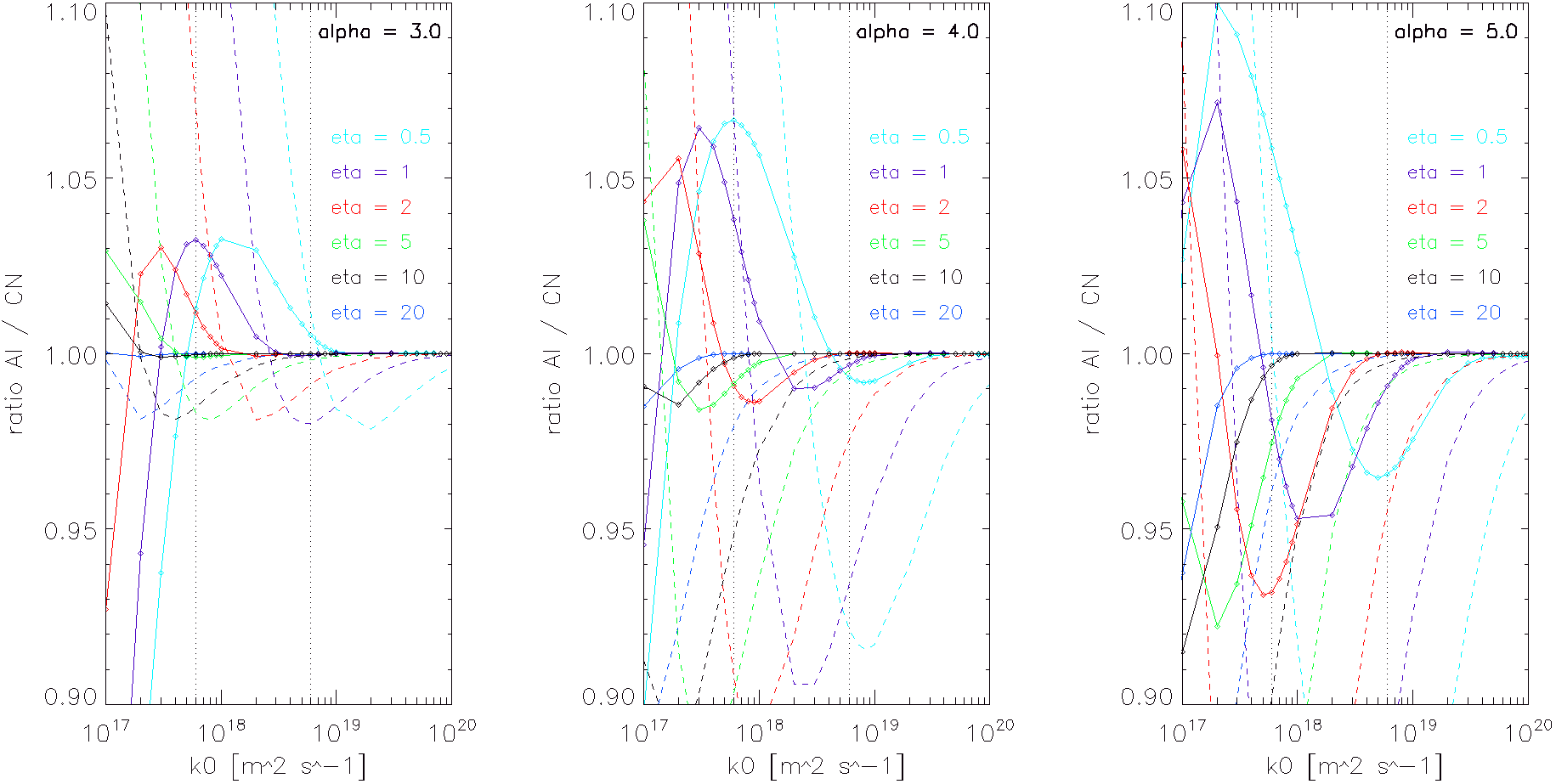}
	\caption{The comparison of the ratio of the AI and CN solutions for different diffusion coefficients $k_0$ and $\alpha=3$ (left-hand panel), 4 (middle panel) and 5 (right-hand panel) at 1 au for $\eta$=0.5, 1, 2, 5, 10, 20. Dashed lines represent the ratio AI$(N_0)$/CN and solid lines represent the ratio AI$(N_0+N_1)$/CN. AI and CN for ${\kappa = k_0\beta\eta}$ (for more details, see the the text)}
	\label{fig8}
\end{figure}

The precision of the AI solution for different values of $k_0$ for ${\kappa = k_0\beta\eta}$ was similarly reviewed, as in the previous two diffusion cases. As one can see from Fig.~\ref{fig8}, the ratio AI/CN goes to unity with increasing values of the two factors, namely with $k_0$ and the energy.
But with increasing $\alpha$, the situation is slightly the opposite and ambiguous. For example, there is about a 3 per cent difference between the AI and CN solutions when $\alpha=3$ for an energy of 111 MeV ($\eta = 0.5$) and $k_0=1\times 10^{18}$\mms, and the difference is already about 5.5 per cent when $\alpha=4$ for the same energy and $k_0$. But when $\alpha=5$, then the difference again becomes 3 per cent for the same parameters. The situation is similar for the other energies. 

For example, when the energy is 1.16 GeV ($\eta = 2$) and $k_0=4\times 10^{17}$\mms, the deferences are 2 per cent for $\alpha=3$, 1 per cent for $\alpha=4$ and 6.5 per cent for $\alpha=5$. Here, it should be noted that if the modulation parameter $x_0$ ($x_0=\frac{ur_0}{k_0}$) is less than 10, i.e. when $k_0>6\times 10^{17}$\mms , then the deviation of the AI solution from the CN solution reaches no more than 6 per cent for any checked $\alpha$ and energy. We think this is a good result taking into account that numerical methods also have accuracy limits, as can be seen from the ratio CN/B-p (Fig.~\ref{fig7}). 
\begin{figure}
	\centering
	\includegraphics[width=.47\textwidth]{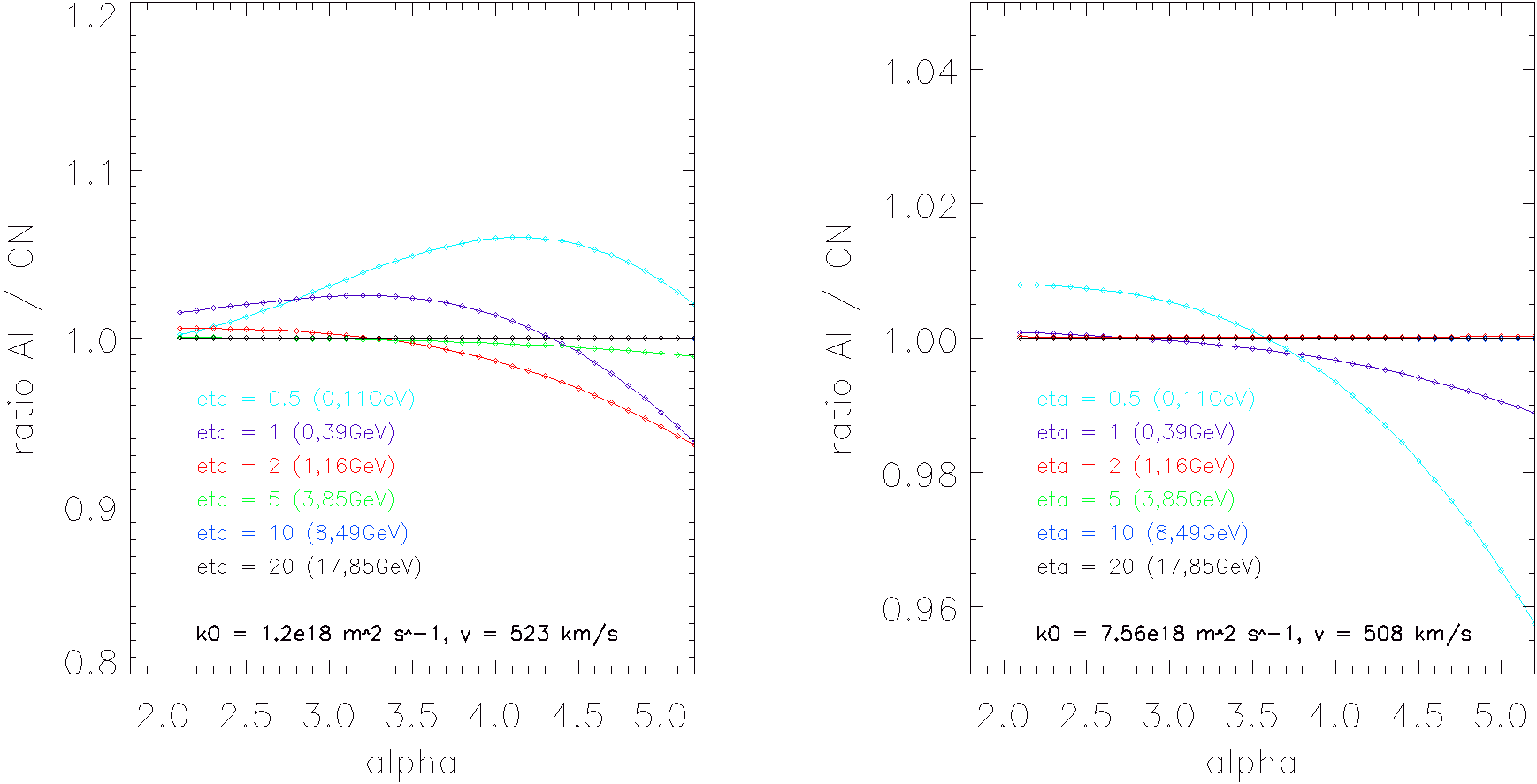}
	\caption{The comparison of the ratio of the AI and CN solutions as functions of the initial spectrum slope ${\alpha}$
		at 1 au for two selected values of ${k_0}$. Solutions for ${\kappa = k_0\beta\eta}$ (for more details, see the text)}
	\label{fig9}
\end{figure}

Finally, the dependence of the precision of the solution on ${\alpha}$ at 1 au for two selected ${k_0}$ is presented in Fig.~\ref{fig9}. The values used for ${k_0}$ were chosen as the border values of the modulation range, between a very strong maximum and a weak minimum of the solar cycle. $k_0$ was evaluated from the modulation strength $\phi$ \citep{Usoskin2005, Usoskin2011}. The weakest modulation during solar cycles 22 and 23 occurred in  1987 February (solar cycle 22) with $\phi=311$ MV. From the average SW speed in 1987 February, $u=5.08\times 10^5$\ms \citep{omniweb}, we found that $k_0=7.56\times 10^{18}$\mms. The strongest modulation during solar cycles 22 and 23 occurred in 1991 June  with $\phi=2016$ MV, $u=5.23\times10^5$\ms , which leads to $k_0=1.2\times 10^{18}$\mms. Using these two values of $k_0$, we can show the precision of the evaluated methods for a range of values appearing in the last solar cycles.
During very strong modulations in solar maxima, when $k_0$ reaches the smallest values, the AI/CN ratios stay within the 1 per cent difference range for energies over 3.85 GeV ($\eta>5$) for all tested values of $\alpha$ (see the left-hand panel in Fig.~\ref{fig9}). The precision is one order of magnitude better during solar minima, when $k_0$ reaches its highest values. As we can see in the right-hand panel of Fig.~\ref{fig9}, the precision is better than 0.1 per cent for energies higher than 1.16 GeV ($\eta>2$) for all tested $\alpha$. It should be noted that the difference is less than the 6 per cent between AI and CN for the solar minima period as well as for the solar maxima period for all tested energies and $\alpha$. 

Let us note that we published codes of models used in this paper on the web page \href{www.CRmodels.org}{www.CRmodels.org}.
\begin{table*}
\centering
\caption{The estimates of the errors (in per cent) of the force-field solutions and AI solutions in comparison with HPF and with the numerical solutions at 1 au for $u=4\times 10^5$\ms, $\alpha=4$ and for various $x_0$ (for more details, see the the text).}
 \label{tab:Tab1}
\begin{tabularx}{\textwidth}{lrrcrcrrrrcrrrr}
\cline{1-15}\\
\multicolumn{0}{c}{}&\multicolumn{4}{c}{$\kappa=$~constant}&\multicolumn{0}{c}{}&\multicolumn{4}{c}{$\kappa=k_0\eta$}&\multicolumn{0}{c}{}&\multicolumn{4}{c}{$\kappa=k_0\beta\eta$}\\[0.2cm]
\cline{2-5}\cline{7-10}\cline{7-10}\cline{12-15}\\
\multicolumn{1}{X}{} & $\frac{N_0}{\rm{HPF}}$ & $\frac{N_0+N_1}{\rm{HPF}}$ & $\frac{N_0+N_1+N_2}{\rm{HPF}}$ & $\frac{\rm{HPF}}{\rm{CN}}$ & & $\frac{N_0}{\rm{CN}}$ & $\frac{N_0+N_1}{\rm{CN}}$ & $\frac{N_0}{\rm{B-p}}$ & $\frac{N_0+N_1}{\rm{B-p}}$ & &$\frac{N_0}{\rm{CN}}$ & $\frac{N_0+N_1}{\rm{CN}}$ & $\frac{N_0}{\rm{B-p}}$ & $\frac{N_0+N_1}{\rm{B-p}}$\\[0.12cm]
\hline \noalign{\vskip 0.1cm}
$x_0=1.2$&7.7&-0.001&-0.009&-0.17& &10.7&1.13&12.16&2.66& &6.58&-0.1&7.77&1.17\\
$x_0=3$&31&3.9&0.08&-0.04& &9.25&0.52&6.54&-1.55& &-2.71&-2.77&0.31&0.26\\
$x_0=5$&55.3&17.1&2.9&0.71& &-2.78&-3.03&3.24&3.24& &-12.1&-5&-8.05&-1.13\\
$x_0=6.5$&69&30.6&8.8&0.96& &-8.77&-4.63&-7.36&-0.74& &-18.2&-5.92&-21.12&-8.47\\
$x_0=9$&83.9&53.6&26.1&1.94& &-21.8&-6.82&-15.01&2.21& &-26.7&-6.57&-43.38&-20.51\\[0.1cm]
\hline
\end{tabularx}
\end{table*}
\begin{table*}
\centering
\caption{The estimates of the errors (in per cent) of the force-field solutions and AI solutions in comparison with the numerical solutions for problems where $\kappa=k_0\eta$ and $\kappa=k_0\beta\eta$ with $x_0=1.2$ ($r=1$ au, $u=4\times 10^5$\ms, $k_0=5\times 10^{18}$\mms), $\alpha=4$ and for energies higher than $\eta=0.2$ ($T=19$ MeV) (for more details, see the the text).}
 \label{tab:Tab2}
\begin{tabularx}{\textwidth}{lXXXXXXXXX}
\cline{1-10}\\
\multicolumn{0}{l}{}&\multicolumn{4}{c}{$\kappa=k_0\eta$}&\multicolumn{0}{c}{}&\multicolumn{4}{c}{$\kappa=k_0\beta\eta$}\\[0.2cm]
\cline{2-5}\cline{7-10}\\
\multicolumn{1}{c}{} & $\frac{N_0}{\rm{CN}}$ & $\frac{N_0+N_1}{\rm{CN}}$ & $\frac{N_0}{\rm{B-p}}$ & $\frac{N_0+N_1}{\rm{B-p}}$ & &$\frac{N_0}{\rm{CN}}$ & $\frac{N_0+N_1}{\rm{CN}}$ & $\frac{N_0}{\rm{B-p}}$ & $\frac{N_0+N_1}{\rm{B-p}}$\\[0.12cm]
 \hline \noalign{\vskip 0.1cm}
$\eta=0.2$ ($T=19$ MeV)&7.36&0.45&8.14&2.08& &-12.73&-5.38&-9.6&-2.45\\
$\eta=0.3$ ($T=41$ MeV)&11.4&1.55&13.23&3.7& &-2.98&-3.13&-1.39&-1.54\\
$\eta=0.4$ ($T=72$ MeV)&11.53&1.44&13.18&3.23& &3.22&-1.23&4.27&-0.12\\
$\eta=0.5$ ($T=111$ MeV)&10.7&1.13&12.16&2.66& &6.58&-0.1&7.77&1.17\\
$\eta=0.6$ ($T=156$ MeV)&9.67&0.84&11.15&2.37& &8.06&0.37&9.01&1.39\\
$\eta=0.7$ ($T=207$ MeV)&8.67&0.61&9.92&1.89& &8.45&0.47&9.53&1.65\\
$\eta=0.8$ ($T=263$ MeV)&7.76&0.44&8.69&1.38& &8.25&0.41&9.13&1.37\\
$\eta=0.9$ ($T=324$ MeV)&6.97&0.32&7.73&1.07& &7.77&0.29&8.51&1.09\\
$\eta=1$ ($T=389$ MeV)&6.28&0.23&6.89&0.84& &7.18&0.17&7.82&0.86\\
$\eta=2$ ($T=1.16$ GeV)&2.68&-0.0008&2.9&0.22& &2.94&-0.18&3.31&0.21\\
$\eta=5$ ($T=3.85$ GeV)&0.64&0.003&0.63&-0.005& &0.59&-0.08&0.71&0.04\\
$\eta=10$ ($T=8.49$ GeV)&0.19&0.006&0.22&0.04& &0.15&-0.04&0.18&-0.01\\
$\eta=20$ ($T=17.85$ GeV)&0.05&0.005&0.08&0.03& &0.03&-0.02&0.06&0.01\\[0.1cm]
\hline
\end{tabularx}
\end{table*}
\section{Discussion and Conclusions}
In this paper, we have obtained solutions for three steady-state problems of CR modulation by means of an AI method. By summing all three cases, we can reach the following conclusion: The accuracy of these solutions at a given point $r$ is dependent more on the value of the modulation parameter $x_0$ ($x_0=\frac{ur_0}{\kappa}$ for the problem, where $\kappa = $~constant and $x_0=\frac{ur_0}{k_0}$ for problems, where $\kappa=k_0\eta$ or $\kappa=k_0\beta\eta$) than on the choice of the slope of the initial spectrum $\alpha$.

Thus, for $x_0<10$, the obtained solutions have minimal differences in accuracy in comparison with numerical solutions and the exact solution for the problem when $\kappa = $~constant. In particular, for the problem where the diffusion coefficient is a constant, so that $x_0=1.2$ (i.e. when $k_0=5 \times 10^{18}$\mms, $u=4\times 10^5$\ms and $r_0=100$ au), the obtained AI solution is even closer to the exact solution than its numerical solution. For a task where the diffusion coefficient is proportional to the momentum, and for the same value of $x_0$, we have shown that the AI solution deviates from the numerical CN solution by not more than $1.55$ per cent over the entire considered range of energies and the deviation is less than $1$ per cent starting from energies of $T=133$ MeV ($\eta=0.55$). For the problem where $\kappa=k_0\beta\eta$, the deviation of the AI solution from the numerical CN solution is less than $1$ per cent beginning at an energy of $T=79$ MeV ($\eta=0.42$), which is better than for the previous problem. It should be noted that although the comparison of the AI solution with the numerical solution B-p is more advantageous for this case (the B-p solution has a deviation of not more than $2.5$ per cent over all of the investigated range of energies) and less profitable for $\kappa=k_0\eta$, here we have provided only one of them because the difference between the two numerical methods can reach $10$ per cent, as seen from Fig.~\ref{fig7}. It should also be noted that the situation with the obtained AI solutions at smaller values of $x_0$, i.e. $x_0<1$, is even better than the above-examined case. 

In this paper, we have evaluated the possible $x_0$ values under real physical conditions; as an example, we have chosen different cycles of solar activity. The maximum $x_0$ value that we have found during maximum solar activity does not exceed $x_0=6.5$ (i.e. when $k_0=1.2\times 10^{18}$\mms, $u=5.23\times 10^5$\ms and $r_0=100$ au). In very rare situations, when there is a very fast SW and a low value of $k_0$, then $x_0$ may reach a value over 10.  

In this case, the difference between the AI and numerical solutions becomes larger but, at the same time, gives us much better accuracy than the force-field solutions ($N_0$). This result is shown in Tables 1 and 2, along with estimates of the errors (in per cent) of $N_0$ and AI solutions in comparison with HPF (only for the problem where $\kappa = $~constant, there is an exact solution) and numerical solutions at 1 au for $u=4\times 10^5$\ms and $\alpha=4$ for various cases. In this table, the case $\kappa = $~constant corresponds to the exact HPF solution only. Therefore, Table~\ref{tab:Tab1} demonstrates estimates for all three problems mentioned above. The parameters were chosen as follows: $x_0=1.2,\:3,\:5,\:6.5,\:9$; energy $\eta=0.5$ ($\kappa=k_0\eta$ and $\kappa=k_0\beta\eta$). Table~\ref{tab:Tab2} displays estimates for the cases when $\kappa$ is not constant (for $x_0=1.2$) and for various energies that are higher than $\eta=0.2$. These tables show the importance of the inclusion of the amendment $N_1$ in comparison with the force-field solution for the selected $x_0$ (Table~\ref{tab:Tab1}) and for the selected $\eta$ (Table~\ref{tab:Tab2}). The result is even better when the second amendment is also taken into account (for the problem with $\kappa = $~constant). We can see that accuracy deteriorates with the increasing $x_0$ (for all the three problems) and with the decreasing $\eta$ (for the problems where $\kappa$ is not constant).

Therefore, we think that using more amendments in this method would produce more accurate solutions for those rare occasions of $x_0$, when it is necessary. To show the above and also that the common iterative solution (\ref{common solution}) converges with the exact solution, a problem was examined in Appendix E, where LIS has the time dependence. For this purpose, a term $\frac{\upartial N_i}{\upartial t}$ in (\ref{iteration eq}) was taken into account. In addition, it has been shown that the exact solution of the considered problem is unknown because of the difficulty to carry out the inverse Laplace transform for equation (\ref{N_eta_s_exact_solution}); therefore, the AI method in this situation, as for another problems, has a big advantage.

So we believe that the obtained solutions for the considered problems by means of the proposed method can be regarded as analytical solutions of such problems with some remarks. The method can be widely applied for solving the various steady-state and non-stationary problems of CR modulation with almost arbitrary dependences of the SW speed, forms of LIS spectrum and the diffusion coefficient ($\kappa$), which is simultaneously dependent on $p$ (or $R$), $\beta$ and $r$. Depending on $u$ and/or $\kappa$, the proposed method may have some restrictions. This is why we have used above the word `almost'. To find at least the zero-order solution of~(\ref{flow r}), we should solve the transcendental equation, which at the same time, depends on $u$, $\kappa$. Usually, it is difficult to solve analytically or even not possible. For example, if $u = $~constant and we choose a form $k$ such as $\kappa=k_0(1+\eta+\eta^2)$, when we insert them into~(\ref{flow r}), we will obtain an equation that may be solved only by a numerical way. Here, it should be noted that `1' in the form of $\kappa$ `prevents' solving this example analytically. We guess that other the restrictions on the use of the method do not exist.

We think that by means of the proposed approach, it is possible to theoretically examine a stationary model of the spherically symmetric heliosphere to describe the physical processes during CR propagation therein \citep{KolesnikShakhov, KotaJokipii2014, Fedorov}. The model includes an environment that consists of two adjacent spherically symmetric (relative to the Sun) regions. The internal part of the heliosphere is limited by a termination shock (TS), which is placed at $r_0$ from the Sun, the radial SW speed is constant and supersonic, $V_{in}$, but $\kappa$ has a form: $\kappa=k_1\beta\eta\frac{r}{r_0}$. Beyond TS, the SW speed jumps to a value $V_{in}/3$ \citep{Richardson2013} and spreads with deceleration in the outer part of the heliosphere (the heliosheath), which is limited by the heliopause (HP). Such deceleration can be presented in the form of a power law depending on the heliocentric distance with index `n', i.e. $V_{ihs}=\frac{V_{in}}{3}{(\frac{r}{r_0})}^{-n}$; here $\kappa$ has the form $\kappa=k_2\beta\eta$, where $k_1$ and $k_2$ may be determined from data obtained by \textit{Voyager 1}. Due to the absence of the SW speed inside the interstellar medium (beyond HP), the phase density of the particles will be determined by only the diffusion process under the constant diffusion coefficient. Note, that \citet{KotaJokipii2014} and \citet{Fedorov} considered the easier mathematical problem, when $n=2$ ($\vect\nabla \cdot \mathbfit {V} = 0$, i.e. beyond the TS, no adiabatic energy changes occur for CRs). It should be noted here that although $\vect\nabla \cdot \mathbfit {V}$ for large $n$ in the heliosheth is a very small quantity, it is not equal to zero. In this case, the third term in (\ref{eq. TPE}) will be proportional to $ \vect\nabla \cdot \mathbfit {V} \cdot \eta \frac{\upartial N}{\upartial \eta}$. This means that neglecting the whole term is not correct, as its contribution may be significant for a range of low-energy particles ($\eta<1$). In a numerical way, this effect was shown by~\citet{Langner2006}, where the symmetrical 2D model for different scenarios, $V_{ihs}$, in the heliosheath was studied. In particular, it was shown there that changing the profiles for $V_{ihs}$ has a noteworthy effect on barrier modulation at low energies and that the barrier at such energies in the heliosheath is wider and deeper for the $V_{ihs} \propto\frac{1}{r^8}$ scenario than for $V_{ihs} \propto\frac{1}{r^2}$. In this way, using this method, we think that it is possible to derive a solution of the problem for given diffusion coefficients as well as with an arbitrary $n$ to look for possible effects on theory.

The method may be applied for the theoretical description of time-dependent CR modulation too, i.e. when we can observe solar-cycle-related changes in CR intensities due to changes in the modulation environment \citep{Manuel2011}. For this purpose, a time-dependent term needs to be taken into account in (\ref{iteration eq}), and changes in the modulation environment will be described by the diffusion coefficient that varies with time.

We also think that the proposed method is a significant step towards developing an analytical approach to solve 2D problems of CR modulation (dependence on the radial distance and polar angle), taking into account particle drift.

\section*{Acknowledgements}

We are grateful to the referee for his/her constructive comments.

This work was supported in part by the National Scholarship Programme of the Slovak Republic for the Support of the Mobility of Students, PhD Students, University Teachers, Researchers and Artists (YLK).

This work was partially supported by the Slovak Academy of Sciences, grant no. MVTS JEM-EUSO.




\begin{thebibliography}{99}
\bibitem[\protect\citeauthoryear{Batalha}{2012}]{Batalha2013}
	Batalha L., 2012, MS thesis, Instituto Superior Tecnico, Universidade Tecnico de Lisboa, Portugal
\bibitem[\protect\citeauthoryear{Bateman $\&$ Erdelyi}{1953}]{Bateman1953}
	Bateman H., Erdelyi A., 1953, Higher Transcendental Functions. Vol. 1.	Available at: \href{http://apps.nrbook.com/bateman/Vol1.pdf}{http://apps.nrbook.com/bateman/Vol1.pdf}
\bibitem[\protect\citeauthoryear{Bobik et al.}{2016}]{Bobik}
	Bobik P. et al., 2016, \jgr, 121, 3920
\bibitem[\protect\citeauthoryear{Cowsik $\&$ Lee}{1977}]{Cowsik}
	Cowsik R., Lee M. A., 1977, \apj, 216, 635
\bibitem[\protect\citeauthoryear{Diaz}{1958}]{Diaz}
	Diaz, J. B., 1958, in Grabbe E. M., Ramo S., Wooldridge D. E., eds, Partial differential equations, in Handbook of Automation, Computation, and Contro. John Wiley, New York, p. 14
\bibitem[\protect\citeauthoryear{Dolginov $\&$ Toptygin}{1967}]{Dolginov1967}
	Dolginov A. Z., Toptygin I. N., 1967, Geomagn. Aeron., 7, 967
\bibitem[\protect\citeauthoryear{Dolginov $\&$ Toptygin}{1968}]{Dolginov1968}
	Dolginov A. Z., Toptygin I. N., 1968, \icarus, 8, 54
\bibitem[\protect\citeauthoryear{Dorman et al.}{1978}]{Dorman1978}
	Dorman L. I., Kats M. E., Fedorov Y. I., Shakhov B. A., 1978, Pis'ma Zh. Ehksp. Teor. Fiz., 27, 374	
\bibitem[\protect\citeauthoryear{Fedorov}{2015}]{Fedorov}
	Fedorov Yu. I., 2015, Kinematics Phys. Celest. Bodies, 31, 105
\bibitem[\protect\citeauthoryear{Fisk}{1971}]{Fisk1971}
	Fisk L. A., 1971, \jgr, 76, 221
\bibitem[\protect\citeauthoryear{Fisk $\&$ Axford}{1969}]{FiskAxford1969}
	Fisk L. A., Axford W. I., 1969, \jgr, 74, 4973 	
\bibitem[\protect\citeauthoryear{Gleeson $\&$ Axford}{1967}]{GleesonAxford1967}
	Gleeson L. J., Axford W. I., 1967, \apj, 149, L115
\bibitem[\protect\citeauthoryear{Gleeson $\&$ Axford}{1968}]{GleesonAxford1968}
	Gleeson L. J., Axford W. I., 1968, \apj, 154, 1011
\bibitem[\protect\citeauthoryear{Gleeson $\&$ Urch}{1973}]{GleesonUrch}
	Gleeson L. J., Urch I. A., 1973, \apss, 25, 387
\bibitem[\protect\citeauthoryear{Gleeson $\&$ Webb}{1978}]{GleesonWebb}
	Gleeson L. J., Webb G. M., 1978, \apss, 58, 21	
\bibitem[\protect\citeauthoryear{Jokipii $\&$ Parker}{1970}]{JokipiiParker}
	Jokipii J. R., Parker E. N., 1970, \apj, 160, 735
\bibitem[\protect\citeauthoryear{Kolesnik $\&$ Shakhov}{2012}]{KolesnikShakhov}
	Kolesnik Yu. L., Shakhov B. A., 2012, Kinematics Phys. Celest. Bodies, 28, 261
\bibitem[\protect\citeauthoryear{Kota}{1977}]{Kota1977}
	Kota J., 1977,  ICRC, 11, 186
\bibitem[\protect\citeauthoryear{Kota $\&$ Jokipii}{2014}]{KotaJokipii2014}
	Kota J., Jokipii J. R., 2014, \apj, 782, 24
\bibitem[\protect\citeauthoryear{Langner et al.}{2006}]{Langner2006}	
	Langner U. W., Potgieter M. S., Fichtner H., Borrmann T., 2006, \apj, 640: 1119
\bibitem[\protect\citeauthoryear{Manuel et al.}{2011}]{Manuel2011}
	Manuel R. S., Ferreira E. S., Potgieter M. S., Strauss R. D., Engelbrecht N. E., 2011, Adv. Space Res., 47, 1529
\bibitem[\protect\citeauthoryear{Morse $\&$ Feshbach}{1953}]{Morse1953}	
	Morse P. M., Feshbach H., 1953, Methods of Theoretical Physics. McGraw-Hill, New York
\bibitem[\protect\citeauthoryear{OMNIweb}{2017}]{omniweb}	
	OMNIweb 2017, Available at: \href{http://omniweb.gsfc.nasa.gov/form/dx1.html}{http://omniweb.gsfc.nasa.gov/form/dx1.html}
\bibitem[\protect\citeauthoryear{Parker}{1965}]{Parker}
	Parker E. N., 1965, \planss, 13, 9
\bibitem[\protect\citeauthoryear{Richardson $\&$ Burlaga}{2013}]{Richardson2013}
	Richardson J. D., Burlaga L. F., 2013, \ssr, 176, 217
\bibitem[\protect\citeauthoryear{Shakhov $\&$ Kolesnik}{2008}]{Shakhov2008}
	Shakhov B. A., Kolesnik Yu. L., 2008, Kinematics Phys. Celest. Bodies, 24, 280	
\bibitem[\protect\citeauthoryear{Shakhov $\&$ Kolesnyk}{2006}]{Shakhov2006}
	Shakhov B. A., Kolesnyk Yu. L., 2006, Kinematika Fiz. Nebesnykh Tel, 22, 101
\bibitem[\protect\citeauthoryear{Shakhov $\&$ Kolesnyk}{2009}]{Shakhov2009}
	Shakhov B. A., Kolesnyk Yu. L., 2009, Proceedings of 21st European Cosmic Ray Symposium, Slovak Academy of Sciences, Ko\v{s}ice, Slovakia, p.245
\bibitem[\protect\citeauthoryear{Usoskin et al.}{2005}]{Usoskin2005}
	Usoskin I. G., Alanko-Huotari K., Kovaltsov G. A., Mursula K., 2005, \jgr, 110, A12108
\bibitem[\protect\citeauthoryear{Usoskin, Bazilevskaya $\&$ Kovaltsov}{2011}]{Usoskin2011}
	Usoskin I. G., Bazilevskaya G. A., Kovaltsov G. A., 2011, \jgr, 116, A02104
\bibitem[\protect\citeauthoryear{Webb $\&$ Gleeson}{1977}]{Webb1977}
	Webb G. M., Gleeson L. J., 1977, \apss, 50, 205
\bibitem[\protect\citeauthoryear{Yamada, Yanagita $\&$ Yoshida}{1998}]{Yamada}
	Yamada Y., Yanagita S., Yoshida T., 1998, \grl, 25, 2353
\bibitem[\protect\citeauthoryear{Zhang}{1999}]{Zhang}
	Zhang M., 1999, \apj, 513, 409	
\end{thebibliography}


\onecolumn
\appendix
\section{ \mbox{\boldmath$\kappa$} = \lowercase{constant}}
To obtain a solution of the problem, we introduce the following dimensionless variables: $x=\frac{ur}{\kappa}, x_0=\frac{ur_0}{\kappa}$ and $\eta=\frac{p}{m_0c}$. Then, taking into account~(\ref{flow r}) and (\ref{first condition}), we obtain a system of equations for finding the zero-order solution:
\begin{equation}
\begin{cases}
\frac{\upartial N_0(x,\:\eta)}{\upartial x}+\frac{\eta}{3}\frac{\upartial N_0(x,\:\eta)}{\upartial \eta}=0,\\
N_0(x_0,\:\eta)=C\eta^{-\alpha}.
\end{cases}
\end{equation}
The last system of equations can be rewritten in the following form:
\begin{equation}
\begin{cases}
N_0(x,\:\eta)=f(\eta\exp(-\frac{x}{3})),\\
N_0(x_0,\:\eta)=C\eta^{-\alpha}.
\end{cases}
\end{equation}
Then, the solution of the last system has the form
\begin{equation}
\label{N_0_k_const}
N_{0}=C\eta^{-\alpha}\exp\bigg[\frac{\alpha}{3}(x-x_{0})\bigg].
\end{equation}
To obtain the first and next amendments, we use a form such as $N_i=C\eta^{-\alpha}\Phi_i(x)$, and when we insert $N_{0}$ into (\ref{iteration eq}), we will obtain an equation for $N_1$:
\begin{equation}
\frac{1}{x^2}\frac{\mathrm{d}}{\mathrm{d}x}x^2\Bigg[-\frac{\mathrm{d}\Phi_1}{\mathrm{d}x}+\frac{\alpha}{3}\Phi_1\Bigg]+\frac{\alpha(3-\alpha)}{9}\exp\bigg[\frac{\alpha}{3}(x-x_{0})\bigg]=0.
\end{equation}
If we expand the exponent in the last equation in a Taylor series, we obtain
\begin{equation}
\frac{\mathrm{d}\Phi_1}{\mathrm{d}x}-\frac{\alpha}{3}\Phi_1=\frac{\alpha(3-\alpha)}{9}\exp\bigg(-\frac{\alpha}{3}x_0\bigg)\sum\limits_{n=0}^\infty\frac{(\frac{\alpha}{3})x^{n+1}}{n!(n+3)}.
\end{equation}
Here, the first condition (\ref{second condition}) was applied, i.e. the integration was carried out from 0 to $x$. We will seek a solution of the last equation in the following form:
\begin{equation}
\Phi_1=B\exp\bigg(\frac{\alpha x}{3}\bigg)+\exp\bigg(\frac{\alpha x}{3}\bigg)\int\limits_0^{x}\exp\bigg(-\frac{\alpha x}{3}\bigg)\frac{\alpha(3-\alpha)}{9}\exp\bigg(-\frac{\alpha}{3}x_0\bigg)\sum\limits_{n=0}^\infty\frac{(\frac{\alpha}{3})x^{n+1}}{n!(n+3)}\mathrm{d}x.
\end{equation}
After an expansion of the exponent in a series followed by integration, we obtain
\begin{equation}
\Phi_1=B\exp\bigg(\frac{\alpha x}{3}\bigg)+\frac{\alpha(3-\alpha)}{9}\exp\bigg(\frac{\alpha}{3}(x-x_0)\bigg)\sum\limits_{n=0}^\infty\sum\limits_{m=0}^\infty\frac{(-1)^m(\frac{\alpha}{3})^{n+m}x^{n+m+2}}{n!m!(n+3)(n+m+2)}.
\end{equation}
The last condition (\ref{second condition}) for finding $B$ was used, i.e. $\Phi_1(x_0)=0$, and then $B$ has the following form:
\begin{equation}
B=-\frac{\alpha(3-\alpha)}{9}\exp\bigg(-\frac{\alpha}{3}x_0\bigg)\sum\limits_{n=0}^\infty\sum\limits_{m=0}^\infty\frac{(-1)^m(\frac{\alpha}{3})^{n+m}x_0^{n+m+2}}{n!m!(n+3)(n+m+2)}.
\end{equation}
And, as a result, the first amendment is
\begin{equation}
N_{1}=C\eta^{-\alpha}\exp\bigg[\frac{\alpha}{3}(x-x_{0})\bigg]\frac{\alpha(3-\alpha)}{9}\sum\limits_{n=0}^\infty\sum\limits_{m=0}^\infty\frac{(-1)^m(\frac{\alpha}{3})^{n+m}(x^{n+m+2}-x_{0}^{n+m+2})}{n!m!(n+m+2)(n+3)}.
\end{equation}
When we perform similar actions in order to find the second amendment, we will obtain\\
\begin{align}
N_{2}=\:&C\eta^{-\alpha}\exp\bigg[\frac{\alpha}{3}(x-x_{0})\bigg]\frac{\alpha(3-\alpha)^2}{27}\sum\limits_{t=0}^\infty\sum\limits_{k=0}^\infty\sum\limits_{n=0}^\infty\sum\limits_{m=0}^\infty\frac{(-1)^{m+t}(\frac{\alpha}{3})^{n+m+k+2}}{t!k!n!m!(n+m+2)(n+3)}\nonumber \\
&\times\Bigg[\frac{n+m+k+2}{(n+m+k+4)(n+m+k+t+3)}(x^{n+m+k+t+3}-x_{0}^{n+m+k+t+3})-\frac{k{x_{0}^{n+m+2}}}{(k+2)(k+t+1)}(x^{k+t+1}-x_{0}^{k+t+1})\Bigg].
\end{align}
An exact solution for this problem was initially obtained by \citet{Dolginov1967}. In our notation, it has the following form:
\begin{equation}
\label{N_ex.sol.}
N_{\mathrm{es}}=C\eta^{-\alpha}\frac{F\bigg(\frac{2\alpha}{3},2;x\bigg)}{F\bigg(\frac{2\alpha}{3},2;x_0\bigg)},
\end{equation}\\
where $F(a,b;x)$ is the confluent hypergeometrical function \citep{Bateman1953}. In particular, from the property of the confluent hypergeometrical function, it follows that for the simpler case when $\alpha=3$, the exact solution~(\ref{N_ex.sol.}) has the form
\begin{equation}
\label{N_ex.sol._alpha_3}
N_{\mathrm{es},\alpha=3}=C\eta^{-3}\exp[x-x_{0}].
\end{equation}

\section{ \mbox{\boldmath$\kappa=\lowercase{k_0}\eta$} }
To obtain a solution for the problem, we introduce the following dimensionless variables: $x=\frac{ur}{k_0}, x_0=\frac{ur_0}{k_0}$ and $\eta=\frac{p}{m_0c}$. Then, taking into account~(\ref{flow r}) and (\ref{first condition}), we obtain a system of questions for finding the zero-order solution:
\begin{equation}
\begin{cases}
\frac{\upartial N_0(x,\:\eta)}{\upartial x}+\frac{1}{3}\frac{\upartial N_0(x,\:\eta)}{\upartial \eta}=0,\\
N_0(x_0,\:\eta)=A\eta^{-\alpha}.
\end{cases}
\end{equation}
The last system can be rewritten in the following form:
\begin{equation}
\begin{cases}
N_0(x,\:\eta)=f(\eta-\frac{x}{3}),\\
N_0(x_0,\:\eta)=A\eta^{-\alpha}.
\end{cases}
\end{equation}
Then, the solution of the last system has the form
\begin{equation}
\label{N_0_eta}
N_{0}=A\left[\eta-\frac{x-x_0}{3}\right]^{-\alpha}.
\end{equation}
After inserting $N_{0}$ into (\ref{iteration eq}), we will obtain an equation for $N_1$:
\begin{equation}
\frac{1}{x^2}\frac{\upartial }{\upartial x}x^2\left[-\eta\frac{\upartial N_1}{\upartial x}-\frac{\eta}{3}\frac{\upartial N_1}{\upartial \eta}\right]=-\frac{\alpha A}{9\eta^2}\frac{\upartial}{\upartial \eta}\left[\eta^3\left(\eta-\frac{x-x_0}{3}\right)^{-\alpha-1} \right].
\end{equation}
After differentiating the left-hand side of the last equation, we obtain
\begin{equation}
-\frac{\upartial N_1}{\upartial x}-\frac{1}{3}\frac{\upartial N_1}{\upartial \eta}=I,
\end{equation}
where
\begin{equation}
I=-\frac{\alpha A}{9\eta x^2}\int\limits_0^{x}
\left(\eta-\frac{\xi-x_0}{3}\right)^{-\alpha-1}\left[3-\eta(\alpha+1)\left(\eta-\frac{\xi-x_0}{3}\right)^{-1} \right]\xi^2\mathrm{d}\xi.
\end{equation}
Here, the limits of integration were selected as being from 0 to $x$ in consequence of applying the first condition~(\ref{second condition}). Now, we consider the integral $I$ separately. If we define $y=\eta-\frac{\xi-x_0}{3}$, then the expression for $I$ can be rewritten in the following form:
\begin{align}
I=&-\frac{\alpha A}{3\eta x^2}\int\limits_{\eta-\frac{x-x_0}{3}}^{\eta+\frac{x_0}{3}}
y^{-\alpha-1}\left[3-\eta(\alpha+1)y^{-1}\right]\left[3\eta-3y+x_0\right]^2\mathrm{d}y=-\frac{\alpha A}{3\eta x^2}\Bigg(\frac{27}{-\alpha+2}\left[\left(\eta+\frac{x_0}{3}\right)^{-\alpha+2}-\left(\eta-\frac{x-x_0}{3}\right)^{-\alpha+2}\right]\nonumber \\
&+\frac{9[\eta(7+\alpha)+2x_0]}{\alpha-1}\left[\left(\eta+\frac{x_0}{3}\right)^{-\alpha+1}-\left(\eta-\frac{x-x_0}{3}\right)^{-\alpha+1}\right]-\frac{3[3\eta+x_0][\eta(5+2\alpha)+x_0]}{\alpha}\left[\left(\eta+\frac{x_0}{3}\right)^{-\alpha}-\left(\eta-\frac{x-x_0}{3}\right)^{-\alpha}\right]\nonumber \\
&+\eta(3\eta+x_0)^2\left[\left(\eta+\frac{x_0}{3}\right)^{-\alpha-1}-\left(\eta-\frac{x-x_0}{3}\right)^{-\alpha-1}\right]\Bigg).
\end{align}
Whereas $N_0$ has the form~(\ref{N_0_eta}), then, taking into account the second condition (\ref{second condition}), it is necessary to search for an $N_1$ of the following form:
\begin{equation}
N_1=\int\limits_x^{x_0}f(x\to\xi,\eta\to\eta-\frac{x-\xi}{3})\mathrm{d}\xi.
\end{equation}
And, as a result, the first amendment takes the following form:
\begin{align}\label{N_1_eta}
N_{1}=&-\frac{\alpha A}{3}\int\limits_x^{x_0}
\frac{27}{-\alpha+2}\frac{\left[\eta-\frac{x-\xi-x_0}{3}\right]^{-\alpha+2}-\left[\eta-\frac{x-x_0}{3}\right]^{-\alpha+2}}{\xi^2\Big(\eta-\frac{x-\xi}{3}\Big)}+\frac{9}{\alpha-1}\frac{\left[\eta-\frac{x-\xi-x_0}{3}\right]^{-\alpha+1}-\left[\eta-\frac{x-x_0}{3}\right]^{-\alpha+1}}{\xi^2\Big(\eta-\frac{x-\xi}{3}\Big)}\nonumber \\
&\times\left[\bigg(\eta-\frac{x-\xi}{3}\bigg)(7+\alpha)+2x_0\right]-\frac{3}{\alpha}\frac{\left[\eta-\frac{x-\xi-x_0}{3}\right]^{-\alpha}-\left[\eta-\frac{x-x_0}{3}\right]^{-\alpha}}{\xi^2(\eta-\frac{x-\xi}{3})}\left[3\bigg(\eta-\frac{x-\xi}{3}\bigg)+x_0\right]\left[\bigg(\eta-\frac{x-\xi}{3}\bigg)(5+2\alpha)+x_0\right]\nonumber \\
&+\frac{\left[\eta-\frac{x-\xi-x_0}{3}\right]^{-\alpha-1}-\left[\eta-\frac{x-x_0}{3}\right]^{-\alpha-1}}{\xi^2}\left[3\bigg(\eta-\frac{x-\xi}{3}\bigg)+x_0\right]^2\:\mathrm{d}\xi\
\end{align}
It should be noted that $N_1$ is not defined for $\alpha=0, 1, 2$.

\section{\mbox{\boldmath$\kappa=\lowercase{k_0}\beta\eta$}}
To obtain a solution for the problem, we introduce the following dimensionless variables: $x=\frac{ur}{k_0}, x_0=\frac{ur_0}{k_0}$ and $\eta=\frac{p}{m_0c}$. Note that for such a problem, the diffusion coefficient $\kappa$ can be rewritten in the following form: $\kappa=k_0\frac{\eta^2}{\sqrt{\eta^2+1}}$. This follows from the fact that $p=\frac{\upupsilon E}{c^2}$ and $E=\sqrt{p^2c^2+m_0^2c^4}$. Taking into account~(\ref{flow r}) and (\ref{first condition}), we obtain a system of equations for finding the zero-order solution:
\begin{equation}
\begin{cases}
\frac{\eta}{\sqrt{\eta^2+1}}\frac{\upartial N_0(x,\:\eta)}{\upartial x}+\frac{1}{3}\frac{\upartial N_0(x,\:\eta)}{\upartial \eta}=0,\\
N_0(x_0,\:\eta)=A\eta^{-\alpha}.
\end{cases}
\end{equation}
The last system can be rewritten in the form
\begin{equation}
\begin{cases}
N_0(x,\:\eta)=f(\sqrt{\eta^2+1}-\frac{x}{3}),\\
N_0(x_0,\:\eta)=A\eta^{-\alpha}.
\end{cases}
\end{equation}
Then, the solution of the last system has the form
\begin{equation}
\label{N_0_beta_eta}
N_{0}=A\left(\sqrt{\left[\sqrt{\eta^2+1}-\frac{x-x_0}{3}\right]^2-1}\right)^{-\alpha}.
\end{equation}
After inserting $N_{0}$ into (\ref{iteration eq}), we will obtain an equation for $N_1$:
\begin{align}
\frac{1}{x^2}&\frac{\upartial }{\upartial x}x^2\left[\frac{\eta^2}{\sqrt{\eta^2+1}}\frac{\upartial N_1}{\upartial x}+\frac{\eta}{3}\frac{\upartial N_1}{\upartial \eta}\right]\nonumber \\
&=-\frac{\alpha A}{3}\left(\left[\sqrt{\eta^2+1}-\frac{x-x_0}{3}\right]^2-1\right)^{-\frac{\alpha}{2}-1}\left[\sqrt{\eta^2+1}-\frac{x-x_0}{3}+\frac{\eta^2}{3\sqrt{\eta^2+1}}\frac{(\alpha+1)\bigg(\sqrt{\eta^2+1}-\frac{x-x_0}{3}\bigg)^2+1}{1-\bigg(\sqrt{\eta^2+1}-\frac{x-x_0}{3}\bigg)^2}\right]
\end{align}
After introducing the variable: $\theta=\sqrt{\eta^2+1}$ and simplifying the last equation, we obtain:
\begin{equation}
\label{N_1_K_L_M}
-\frac{\upartial N_1}{\upartial x}-\frac{1}{3}\frac{\upartial N_1}{\upartial \theta}=-\frac{\alpha A}{3x^2 (\theta^2-1)}\left[ \left(\frac{4\theta^2-1}{3}+\frac{x_0\theta}{3}\right)K-\frac{\theta}{3}L-\frac{(\alpha+2)(\theta^2-1)}{3}M \right],
\end{equation}
where
\\$K=\int\limits_0^{x}\left(\left[\theta-\frac{x-x_0}{3}\right]^2-1\right)^{-\frac{\alpha}{2}-1} x^2\mathrm{d}x;$
\\$L=\int\limits_0^{x}\left(\left[\theta-\frac{x-x_0}{3}\right]^2-1\right)^{-\frac{\alpha}{2}-1} x^3\mathrm{d}x;$
\\$M=\int\limits_0^{x}\left(\left[\theta-\frac{x-x_0}{3}\right]^2-1\right)^{-\frac{\alpha}{2}-2}\left[\theta-\frac{x-x_0}{3}\right]^2 x^2\mathrm{d}x$.

Note that the limits of integration in $K$, $L$, $M$ were selected as being from 0 to $x$ in consequence of applying of the first condition~(\ref{second condition}). The detailed calculation of the integrals was explained in Appendix D.

Due to the fact that the solution of the homogeneous equation~(\ref{N_1_K_L_M}) has the form~$f(\theta-\frac{x}{3})$, and taking into account the second condition (\ref{second condition}), it is necessary to search for an $N_1$ in the following form:
\begin{equation}
\label{N_1_form_beta_eta}
N_1=\int\limits_x^{x_0}f\Bigg (x\to\xi,\theta\to\theta-\frac{x-\xi }{3}\Bigg) \mathrm{d}\xi.
\end{equation}
By taking into account the form for $N_1$~(\ref{N_1_form_beta_eta}) and also using~(\ref{N_1_K_L_M}), we obtain
\begin{align}\label{N_1_beta_eta}
N_1=&-\frac{\alpha A}{9}\int\limits_x^{x_0}\frac{\mathrm{d}\xi}{\xi^2\left((\theta-\frac{x-\xi }{3})^2-1\right)} \Bigg[\left(4\bigg(\theta-\frac{x-\xi }{3}\bigg)^2
+x_0\bigg(\theta-\frac{x-\xi }{3}\bigg)-1\right)K \bigg(x\to\xi,\theta\to\theta-\frac{x-\xi }{3}\bigg)-\bigg(\theta-\frac{x-\xi }{3}\bigg)\nonumber \\ 
&\times L\bigg(x\to\xi,\theta\to\theta-\frac{x-\xi }{3}\bigg)-(\alpha+2) \bigg((\theta-\frac{x-\xi }{3})^2-1\bigg) M\bigg(x\to\xi,\theta\to\theta-\frac{x-\xi }{3}\bigg) \Bigg].
\end{align}

\section{calculation of the integrals $\mathbfit K$, $\mathbfit L$, $\mathbfit M$}
For the calculation of the integrals $K$, $L$, $M$ from Appendix C, let us first consider the integral $I$ that has the following form:
\begin{equation}
I=\int z^k(z^2-1)^\beta \mathrm{d}z.
\end{equation}
If we represent $(z^2-1)^\beta$ as a sum, i.e. $(z^2-1)^\beta=\sum\limits_{n=0}^\infty (-1)^n C_\beta^n\cdot (z^2)^{\beta-n}$, where $C_\beta^n$ is the binomial coefficient (which can be represented through the Gamma function $\Gamma$ as $C_\beta^n=\frac{\Gamma(\beta+1)}{\Gamma(\beta-n+1) n!}$), we can apply here properties of $\Gamma$ such as $\frac{\Gamma(\beta+1)}{\Gamma(\beta-n+1)}=\frac{\Gamma(n-\beta)}{\Gamma(-\beta)}(-1)^n$
and see that $I$ assumes the following form:
\begin{align}
\label{spec_integral}
I=-\frac{z^{k+2\beta+1}}{2}\sum\limits_{n=0}^\infty \frac{\Gamma(n-\beta)}{\Gamma(-\beta) n!(n-\beta-\frac{k}{2}-\frac{1}{2})}z^{-2n}.
\end{align}\\
If we now use the fact that $\Gamma(z)=z\Gamma(z+1)$ and multiply~(\ref{spec_integral}) by $\frac{\Gamma(-\frac{k}{2}+\frac{1}{2}-\beta)}{\Gamma(-\frac{k}{2}-\frac{1}{2}-\beta)}$, then we obtain
\begin{equation}
I=-\frac{z^{k+2\beta+1}}{2}\frac{\Gamma(-\frac{k}{2}-\frac{1}{2}-\beta)}{\Gamma(-\frac{k}{2}+\frac{1}{2}-\beta)}
\sum\limits_{n=0}^\infty \frac{\Gamma(-\beta+n)\Gamma(n-\frac{k}{2}-\frac{1}{2}-\beta)\Gamma(-\frac{k}{2}+\frac{1}{2}-\beta)}{\Gamma(-\beta)n!\Gamma(n-\frac{k}{2}+\frac{1}{2}-\beta)\Gamma(-\frac{k}{2}-\frac{1}{2}-\beta)}z^{-2n}.
\end{equation}
For the last expression, we can apply the definition of hypergeometrical function $F(a,b;c;z)$ \citep{Bateman1953}:
$F(a,b;c;z)=\sum\limits_{n=0}^\infty \frac{\Gamma(a+n)}{\Gamma(a)}\frac{\Gamma(b+n)}{\Gamma(b)}\frac{\Gamma(c)}{\Gamma(c+n)}\frac{z^n}{n!}$. Finally, we obtain
\begin{equation}
\label{Integral_I}
\int z^k(z^2-1)^\beta \mathrm{d}z=\frac{z^{k+1+2\beta}}{k+1+2\beta}F(-\beta,-\frac{k}{2}-\frac{1}{2}-\beta;-\frac{k}{2}+\frac{1}{2}-\beta;z^{-2}).
\end{equation}
And now, to use (\ref{Integral_I}) to find the integrals $K$, $L$, $M$, we substitute $\theta-\frac{x-x_0}{3}=z$ in these integrals. As a result, these integrals take the following form:
\\$K=27\int\limits_{\theta-\frac{x-x_0}{3}}^{\theta+\frac{x_0}{3}}\left(z^2-1\right)^{-\frac{\alpha}{2}-1}(z^2-2\omega z+\omega^2)\mathrm{d}z;$
\\$L=-81\int\limits_{\theta-\frac{x-x_0}{3}}^{\theta+\frac{x_0}{3}}\left(z^2-1\right)^{-\frac{\alpha}{2}-1}(z^3-3\omega z^2+3\omega^2z-\omega^3)\mathrm{d}z;$
\\$M=27\int\limits_{\theta-\frac{x-x_0}{3}}^{\theta+\frac{x_0}{3}}\left(z^2-1\right)^{-\frac{\alpha}{2}-2}(z^4-2\omega z^3+\omega^2z^2)\mathrm{d}z$,\\
where $\omega=\theta+\frac{x_0}{3}$. Eventually, if we apply (\ref{Integral_I}) to each of these integrals, we
obtain
\begin{align}
K=&\Bigg [\frac{27}{1-\alpha}z^{1-\alpha} F\bigg(\frac{\alpha}{2}+1,\frac{\alpha}{2}-\frac{1}{2};\frac{\alpha}{2}+\frac{1}{2};\frac{1}{z^2}\bigg)+\frac{54}{\alpha}\omega z^{-\alpha} F\bigg(\frac{\alpha}{2}+1,\frac{\alpha}{2};\frac{\alpha}{2}+1;\frac{1}{z^2}\bigg)\nonumber\\
&-\frac{27}{\alpha+1}\omega^2 z^{-\alpha-1}F\bigg(\frac{\alpha}{2}+1,\frac{\alpha}{2}+\frac{1}{2};\frac{\alpha}{2}+\frac{3}{2};\frac{1}{z^2}\bigg)\Bigg ]_{\theta-\frac{x-x0}{3}}^{\theta+\frac{x_0}{3}},
\end{align}
\begin{align}
L=&\Bigg [\frac{81}{\alpha-2}z^{2-\alpha} F\bigg(\frac{\alpha}{2}+1,\frac{\alpha}{2}-1;\frac{\alpha}{2};\frac{1}{z^2}\bigg)+\frac{243}{1-\alpha}\omega z^{1-\alpha}F\bigg(\frac{\alpha}{2}+1,\frac{\alpha}{2}-\frac{1}{2};\frac{\alpha}{2}+\frac{1}{2};\frac{1}{z^2}\bigg)\nonumber\\
&+\frac{243}{\alpha}\omega^2 z^{-\alpha} F\bigg(\frac{\alpha}{2}+1,\frac{\alpha}{2};\frac{\alpha}{2}+1;\frac{1}{z^2}\bigg)-\frac{81}{\alpha+1}\omega^3 z^{-\alpha-1}F\bigg(\frac{\alpha}{2}+1,\frac{\alpha}{2}+\frac{1}{2};\frac{\alpha}{2}+\frac{3}{2};\frac{1}{z^2}\bigg)\Bigg ]_{\theta-\frac{x-x_0}{3}}^{\theta+\frac{x_0}{3}},
\end{align}
\begin{align}
M =&\Bigg [\frac{27}{-\alpha+1}z^{-\alpha+1} F\bigg(\frac{\alpha}{2}+2,\frac{\alpha}{2}-\frac{1}{2};\frac{\alpha}{2}+\frac{1}{2};\frac{1}{z^2}\bigg)+\frac{54}{\alpha}\omega z^{-\alpha} F\bigg(\frac{\alpha}{2}+2,\frac{\alpha}{2};\frac{\alpha}{2}+1;\frac{1}{z^2}\bigg)\nonumber\\
&-\frac{27}{\alpha+1}\omega^2 z^{-\alpha-1} F\bigg(\frac{\alpha}{2}+2,\frac{\alpha}{2}+\frac{1}{2};\frac{\alpha}{2}+\frac{3}{2};\frac{1}{z^2}\bigg)\Bigg ]_{\theta-\frac{x-x_0}{3}}^{\theta+\frac{x_0}{3}}.
\end{align}
\section{\mbox{\boldmath$\kappa$} = \lowercase{constant} and LIS has the time dependence}

Let us consider a problem when the diffusion coefficient ($\kappa$) is constant inside the heliosphere, but LIS has a form $N_0(r_0,\:\eta,\:t)=f(t)\eta^{-\alpha}$, i.e. LIS has dependence on the time and momentum of the particle on the boundary of the heliosphere $r_0$. For example, such dependence may be considered during a supernova explosion, when the density of the particles on the border of the heliosphere varies with time. To obtain solutions, i.e. as an exact solution as well as by means of the method, we introduce the following dimensionless variables: $x=\frac{ur}{\kappa}, x_0=\frac{ur_0}{\kappa}$, $\eta=\frac{p}{m_0c}$ and $ \tau=\frac{u^2t}{\kappa}$. 

The exact solution of the problem may be obtained if to solve a system of equations:
\begin{align*}
\begin{cases}
\frac{1}{x^2}\frac{\upartial }{\upartial x}x^2\left[-\frac{\upartial N(x,\:\eta,\:\tau)}{\upartial x}-\frac{\eta}{3}\frac{\upartial N(x,\:\eta,\:\tau)}{\upartial \eta}\right]+\frac{1}{3\eta^2}\frac{\upartial}{\upartial \eta}\left[\eta^3\frac{\upartial N(x,\:\eta,\:\tau)}{\upartial x} \right]=-\frac{\upartial N(x,\:\eta,\:\tau)}{\upartial \tau},\\
N(x_0,\:\eta,\:\tau)=f(\tau)\eta^{-\alpha}.
\end{cases}
\end{align*}

Here, the first equation is TPE in a form (\ref{eq. TPE}), taking into account the dimensionless variables, and the second equation is the boundary condition. Applying the Laplace transform, i.e. $N(x,\:\eta,\:s)=\int\limits_{0}^{\infty}\exp(-s\tau)N(x,\:\eta,\:\tau)\mathrm{d}\tau$ to the last system, we obtain
\begin{align}
\begin{cases}
\frac{\upartial^2 N(x,\:\eta,\:s)}{\upartial x^2}+ (\frac{2}{x}-1)\frac{\upartial N(x,\:\eta,\:s)}{\upartial x}+\frac{2\eta}{3x}\frac{\upartial N(x,\:\eta,\:s)}{\upartial \eta}=sN(x,\:\eta,\:s),\\
N(x_0,\:\eta,\:s)=F(s)\eta^{-\alpha}.
\end{cases}
\end{align}

Let us note that during integrating of the right-hand side of the last system, we used the fact that there was no phase density at the initial time, i.e. $N(x,\:\eta,\:0)=0$. The solution of the last system has the following form:
\begin{equation}
\label{N_eta_s_exact_solution}
N(x,\:\eta,\:s)=F(s)\eta^{-\alpha}\exp\bigg[-\frac{1}{2}(x-x_0)(\sqrt{4s+1}-1)\bigg]\frac{F\bigg(1+\frac{2\alpha-3}{3\sqrt{4s+1}},2;\sqrt{4s+1}x\bigg)}{F\bigg(1+\frac{2\alpha-3}{3\sqrt{4s+1}},2;\sqrt{4s+1}x_0\bigg)}
\end{equation}

If we expand the exponent and the confluent hypergeometrical function in the last equation in a Taylor series and restrict ourselves by the second-order $x$ and $x_0$ in such expansion, we obtain
\begin{align}\label{N_eta_s_ex.sol.}
N(x,\:\eta,\:s) & =F(s)\eta^{-\alpha}\frac{1+\Big(1-\sqrt{4s+1}\Big)\frac{x}{2}+\Big(1-\sqrt{4s+1}\Big)^2\frac{x^2}{8}+...}{1+\Big(1-\sqrt{4s+1}\Big)\frac{x_0}{2}+\Big(1-\sqrt{4s+1}\Big)^2\frac{x_0^2}{8}+...}\frac{1+\frac{3\sqrt{4s+1}+2\alpha-3}{6}x+\frac{(3\sqrt{4s+1}+2\alpha-3)(6\sqrt{4s+1}+2\alpha-3)}{108}x^2+...}{1+\frac{3\sqrt{4s+1}+2\alpha-3}{6}x_0+\frac{(3\sqrt{4s+1}+2\alpha-3)(6\sqrt{4s+1}+2\alpha-3)}{108}x_0^2+...} \nonumber\\
& =F(s)\eta^{-\alpha}\frac{1+\frac{\alpha}{3}x+\Big(\frac{1}{6}s+\frac{1}{18}\alpha+\frac{1}{27}\alpha^2\Big)x^2+...}{1+\frac{\alpha}{3}x_0+\Big(\frac{1}{6}s+\frac{1}{18}\alpha+\frac{1}{27}\alpha^2\Big)x_0^2+...}.
\end{align}

Now, divide the numerator of (\ref{N_eta_s_ex.sol.}) ($1+\frac{\alpha}{3}x+(\frac{1}{6}s+\frac{1}{18}\alpha+\frac{1}{27}\alpha^2)x^2+...$) by the 
denominator of (\ref{N_eta_s_ex.sol.}) ($1+\frac{\alpha}{3}x_0+(\frac{1}{6}s+\frac{1}{18}\alpha+\frac{1}{27}\alpha^2)x_0^2+...$) in a column

$$\begin{array}{r@{\,}l|l}
\:\:\:\:\:\:\:\:\dropsign{-} 1+\frac{\alpha}{3}x+(\frac{1}{6}s+\frac{1}{18}\alpha+\frac{1}{27}\alpha^2)x^2\phantom{000000000000000000000}&\phantom{00}&\,\rule{0cm}{0.5cm} 1+\frac{\alpha}{3}x_0+(\frac{1}{6}s+\frac{1}{18}\alpha+\frac{1}{27}\alpha^2)x_0^2 \\ \cline{3-3}\rule{0cm}{0.5cm}
1+\frac{\alpha}{3}x_0+(\frac{1}{6}s+\frac{1}{18}\alpha+\frac{1}{27}\alpha^2)x_0^2\phantom{00000000000000000000}&&\,1+\frac{\alpha}{3}(x-x_0)+(\frac{1}{6}s+\frac{1}{18}\alpha)(x^2-x_0^2)-\frac{\alpha^2}{9}x_0x+\frac{\alpha^2}{27}(x^2+2x_0^2) \\ \cline{1-1} \\[\dimexpr-\normalbaselineskip+\jot]
\dropsign{-} \frac{\alpha}{3}(x-x_0) +(\frac{1}{6}s+\frac{1}{18}\alpha+\frac{1}{27}\alpha^2)(x^2-x_0^2)\phantom{........} \\\rule{0cm}{0.5cm}
\frac{\alpha}{3}(x-x_0) +\frac{\alpha^2}{9}x_0(x-x_0)+...\phantom{.......................}\\ \cline{1-1} \\[\dimexpr-\normalbaselineskip+\jot]
\:\:\:\:\:\:\:\:\:\dropsign{-} (\frac{1}{6}s+\frac{1}{18}\alpha)(x^2-x_0^2)-\frac{\alpha^2}{9}x_0x+\frac{\alpha^2}{27}(x^2+2x_0^2)\\\rule{0cm}{0.5cm}
(\frac{1}{6}s+\frac{1}{18}\alpha)(x^2-x_0^2)-\frac{\alpha^2}{9}x_0x+\frac{\alpha^2}{27}(x^2+2x_0^2)\\ 
\cline{1-1}\rule{0cm}{0.4cm} 
0+...
\end{array}
$$

And as a result, the analytical image of the problem retaining until the second order of $x$ and $x_0$ has the following form:
\begin{equation}
\label{N_eta_s}
N(x,\:\eta,\:s)=F(s)\eta^{-\alpha}\Bigg[1+\frac{\alpha}{3}(x-x_0)-\frac{\alpha^2}{9}x_0x+\bigg(\frac{1}{6}s+\frac{1}{18}\alpha\bigg)(x^2-x_0^2)+ \frac{\alpha^2}{27}(x^2+2x_0^2)\Bigg].
\end{equation}

To obtain the analytical solution for the problem, it is necessary to apply the inverse Laplace transform, i.e. $f(\tau)=\frac{1}{2\upi i}\int\limits_{\sigma_1-i\infty}^{\sigma_1+i\infty}\exp(s\tau)F(s)\mathrm{d}s$. Note that in contrast to (\ref{N_eta_s_exact_solution}), this approach can be readily applied for the last equation, subject to the restriction that has been imposed above. As a result, we obtain an analytic solution for the problem:
\begin{align}
\label{N_eta_tau}
N(x,\:\eta,\:\tau)=\eta^{-\alpha} \Bigg[ 1+\frac{\alpha}{3}(x-x_0)-\frac{\alpha^2}{9}x_0x+\frac{1}{18}\alpha(x^2-x_0^2)+\frac{\alpha^2}{27}(x^2+2x_0^2) \Bigg ] f(\tau)+ \eta^{-\alpha}\Bigg[\frac{x^2-x_0^2}{6}\Bigg]\frac{\upartial f(\tau)}{\upartial \tau}.
\end{align}
Now we try to find a solution to the problem by the AI method. Taking into account~(\ref{flow r}) and (\ref{first condition}), we obtain a system of equations for finding the zero-order solution:
\begin{equation}
\begin{cases}
\frac{\upartial N_0(x,\:\eta,\:\tau)}{\upartial x}+\frac{\eta}{3}\frac{\upartial N_0(x,\:\eta,\:\tau)}{\upartial \eta}=0,\\
N_0(x_0,\:\eta,\:\tau)=f(\tau)\eta^{-\alpha}.
\end{cases}
\end{equation}
Then, the solution of the last has the form:
\begin{equation}
\label{N_0_eta_time}
N_{0}(x,\:\eta,\:s)=F(s)\eta^{-\alpha}\exp\bigg[\frac{\alpha}{3}(x-x_{0})\bigg].
\end{equation}
Note that the image of the zero-order solution~(\ref{N_0_eta_time}) has a similar form to~(\ref{N_0_k_const}) for a problem, which was detailed in Appendix A. To obtain the first amendment, we use a form such as $N_1=\eta^{-\alpha}\Phi_1(x)$, and when we insert $N_{0}$ into (\ref{iteration eq}), we obtain an equation for the image of $N_1$:
\begin{equation}
\frac{1}{x^2}\frac{\mathrm{d}}{\mathrm{d}x}x^2\left[-\frac{\mathrm{d}\Phi_1}{\mathrm{d}x}+\frac{\alpha}{3}\Phi_1\right]+\frac{\alpha(3-\alpha)}{9}\exp\bigg[\frac{\alpha}{3}(x-x_{0})\bigg]=-sF(s)\exp\bigg[\frac{\alpha}{3}(x-x_{0})\bigg].
\end{equation} 
When we perform the same actions as were done in Appendix A  while the obtaining $N_1$, we finally obtain
\begin{align}
\label{N_1_eta_time}
N_{1}(x,\:\eta,\:s)=F(s)\eta^{-\alpha}\bigg[\frac{\alpha(3-\alpha)}{9}+s\bigg]\exp\bigg[\frac{\alpha}{3}(x-x_{0})\bigg]\sum\limits_{n=0}^\infty\sum\limits_{m=0}^\infty\frac{(-1)^m(\frac{\alpha}{3})^{n+m}(x^{n+m+2}-x_{0}^{n+m+2})}{n!m!(n+m+2)(n+3)}.
\end{align}
If we expand the exponents in a series after retaining until the second order of $x$ and $x_0$ as it has been done in finding analytical solution, we obtain
\begin{align}\label{N_0+1_eta_s}
N_{0}(x,\:\eta,\:s)+N_{1}(x,\:\eta,\:s) & =F(s)\eta^{-\alpha}\Bigg[1+\frac{\alpha}{3}(x-x_0)+\frac{\alpha^2}{18}(x^2+x_0^2)-\frac{\alpha^2}{9}xx_0\Bigg]+F(s)\eta^{-\alpha}\Bigg[\frac{\alpha}{18}(x^2-x_0^2)-\frac{\alpha^2}{54}(x^2-x_0^2)+\frac{s}{6}(x^2-x_0^2) \Bigg] \nonumber\\
& = F(s)\eta^{-\alpha} \Bigg[1+\frac{\alpha}{3}(x-x_0)-\frac{\alpha^2}{9}x_0x+(\frac{s}{6}+\frac{\alpha}{18})(x^2-x_0^2)+\frac{\alpha^2}{27}(x^2+2x_0^2)\Bigg].
\end{align}

It should be noted that here we restricted ourselves to only the first amendment because the following amendments will contain $x$ and $x_0$ higher than the second order, as it can be seen in Appendix A. 

It can be seen that every term of the image of the AI solution (\ref{N_0+1_eta_s}) coincides with the corresponding term of the analytical image (\ref{N_eta_s}). Of course, own solutions of these images are identical to each other, i.e. (\ref{N_eta_tau}) is a solution for both (\ref{N_0+1_eta_s}) and (\ref{N_eta_s}). Furthermore, if we examine our problem till the first order $x$ and $x_0$, we see that the first three terms of the image (\ref{N_0+1_eta_s}) (that included in the $N_0$) coincide with the first three terms of image (\ref{N_eta_s}). This indicates that the common iterative solution (\ref{common solution}) corresponds and converges with the exact solution. In other words, using more amendments by means of the AI method would bring it closer to the exact solution of the considered problem.
\bsp	
\label{lastpage}
\end{document}